\documentclass[a4paper]{article}
\usepackage{graphicx}
\usepackage{morefloats} 
\usepackage{color}
\usepackage{amsmath}
\usepackage{amsfonts}
\usepackage{amssymb}
\usepackage[colorinlistoftodos]{todonotes}
\usepackage{algorithm}
\usepackage{algorithmic}
\usepackage{fix-cm}
\usepackage[normalem]{ulem}
\usepackage{a4wide}

\def\bfx{\mbox{\boldmath$x$}}
\def\bfv{\mbox{\boldmath$v$}}

\newtheorem{remark}{Remark}
\newtheorem{example}{Example}

\title{Numerical extraction of a macroscopic PDE
  and a lifting operator from a Lattice Boltzmann model}

\author{Ynte Vanderhoydonc\thanks{Dept.~Mathematics and Computer Science, Universiteit Antwerpen, Middelheimlaan 1, 2020 Antwerpen, Belgium ({\tt ynte.vanderhoydonc@ua.ac.be}).}
        \and
        Wim Vanroose\thanks{Dept.~Mathematics and Computer Science, Universiteit Antwerpen, Middelheimlaan 1, 2020 Antwerpen, Belgium ({\tt wim.vanroose@ua.ac.be}).}}

\date{}

\begin{document}

\maketitle

%
%

\begin{abstract}
  Lifting operators play an important role in starting a lattice
  Boltzmann model from a given initial density. The density, a
  macroscopic variable, needs to be mapped to the distribution
  functions, mesoscopic variables, of the lattice Boltzmann
  model. Several methods proposed as lifting operators have been
  tested and discussed in the literature. The most famous methods are
  an analytically found lifting operator, like the Chapman-Enskog
  expansion, and a numerical method, like the Constrained Runs
  algorithm, to arrive at an implicit expression for the unknown
  distribution functions with the help of the density. This paper
  proposes a lifting operator that alleviates several drawbacks of
  these existing methods. In particular, we focus on the computational
  expense and the analytical work that needs to be done. The proposed
  lifting operator, a numerical Chapman-Enskog expansion, obtains the
  coefficients of the Chapman-Enskog expansion numerically. Another
  important feature of the use of lifting operators is found in hybrid
  models. There the lattice Boltzmann model is spatially coupled with
  a model based on a more macroscopic description, for example an
  advection-diffusion-reaction equation. In one part of the domain,
  the lattice Boltzmann model is used, while in another part, the more
  macroscopic model. Such a hybrid coupling results in missing data at
  the interfaces between the different models. A lifting operator is
  then an important tool since the lattice Boltzmann model is
  typically described by more variables than a model based on a
  macroscopic partial differential equation.
\end{abstract}

%
%

\section{Introduction}

A lifting operator is, in a multiscale method, an important tool that
maps macroscopic variables to microscopic/mesoscopic variables.  In
kinetic models, for example, a lifting operator will map low order
moments, like the density $\rho(\bfx,t)$ that counts the number of
particles in a point $\bfx \in D_{\bfx} \subset \mathbb{R}^n$, $n \in
\mathbb{N}_0$ to a distribution function $f(\bfx,\bfv,t)$ that counts
the number of particles in a point $(\bfx, \bfv)$ in phase space where
$\bfx \in D_{\bfx} \subset \mathbb{R}^n$ and the velocity $\bfv \in
D_{\bfv} \subset \mathbb{R}^n$. For practical applications one uses
$n\in\{1,2,3\}$ and time $t \geq 0$.

In these problems the macroscopic level is typically described by a
few low order moments and their evolution is simulated by use of a
macroscopic partial differential equation (PDE).  For example, the
evolution of the density $\rho(\bfx,t)$ can be represented by an
advection-diffusion-reaction equation. While the
microscopic/mesoscopic level is typically described by a
Boltzmann equation that evolves the distribution function
$f(\bfx,\bfv,t)$.  A lattice Boltzmann model (LBM) is a special
discretization of the Boltzmann equation that is, for example, used to
simulate complex fluid systems.

Some examples of complex flows for which LBMs are used are flows in
complicated geometries, multiphase and turbulent flows. Applications
can be found in \cite{benzi_succi, succi} and a more recent review is
given by Aidun \textit{et al.}~\cite{aidun}. The application of the
lattice Boltzmann model to multiscale physics in fluids is discussed
in \cite{succi_filippova}. Banda \textit{et al.}~\cite{banda} present 
a high order relaxation system for the multiscale lattice Boltzmann equation 
to obtain the incompressible Navier-Stokes limit.


%
Kevrekidis \textit{et al.}~introduced a lifting operator to couple
different scales in a dynamical system in the equation-free framework
\cite{kevrekidis}. This allows to model the dynamics at the
macroscopic level by using short bursts of the microscopic simulation.

A problem of lattice Boltzmann methods is the determination of initial conditions, 
usually given by macroscopic variables.
During initialization and spatial coupling in a hybrid model, 
a one-to-many map needs to be created, known as a lifting operator. In this article we discuss a
hybrid LBM and PDE model that uses a lattice Boltzmann model in one
part of the domain while another part of the domain is described by a
macroscopic partial differential equation. These different levels of
description create missing data at the interfaces that can be resolved
with a lifting operator.

Hybrid approaches have been formulated for various flow problems. A
Lennard-Jones particle dynamics is coupled with a compressible
Navier-Stokes in \cite{hybrid}.  A Boltzmann, respectively a lattice
Boltzmann, model is coupled to the Navier-Stokes equations in
\cite{leTallec} and \cite{latt}.  A LBM is also coupled with a
Navier-Stokes in Peano, an adaptive mesh refinement framework with
spacetree grids \cite{peano}. Furthermore, the Boltzmann equation is
coupled to the Euler equations in \cite{bourgat}. 

Coupled models also play an important role in the simulation of
materials. A review of atomistic-to-continuum coupling is found in
\cite{atomistic_to_continuum}. A more detailed overview for coupling
methods in hybrid models can be found in \cite{garcia} and
\cite{koumoutsakos}.  In \cite{garcia} adaptive mesh and algorithm
refinement is used in parts of the domain where a continuum
description is replaced by a particle description. Coupled molecular
dynamics and lattice Boltzmann models based on Schwarz's alternating
method is presented in \cite{koumoutsakos}. Dimarco \textit{et al.}~\cite{dimarco} 
merge deterministic methods for the equilibrium part
with particle methods for the nonequilibrium part and present results
for the Boltzmann equation with Bhatnagar-Gross-Krook (BGK) approximation.

 The coupling --- that will be discussed in this article --- of
 LBMs with reaction-diffusion PDEs is earlier considered in \cite{leemput, leemput_phd, vanderhoydonc}. 

 We propose a general lifting operator that maps densities to
 distribution functions. It is illustrated for a LBM, but we believe that it is
 applicable to general discretizations of the Boltzmann equation and 
 it can map more moments to the corresponding distribution functions.

 The new method will be compared to the Chapman-Enskog expansion
 \cite{chapman_cowling}, a well known analytical method for the
 initialization of a lattice Boltzmann model, and the Constrained Runs
 (CR) algorithm \cite{initialization}, a numerical lifting
 operator. The Chapman-Enskog expansion writes the distribution
 functions as an analytical series of the density. The Constrained
 Runs algorithm is based on the attraction of the dynamics toward the
 slow manifold and expresses, in an implicit way, the unknown
 distribution functions with the help of the density in successive
 grid points.  A numerical comparison of these methods is given in
 \cite{vanderhoydonc} for hybrid models that spatially couple a
 diffusion PDE model and a LBM. 
 Although the Constrained Runs algorithm is very accurate, its major
 drawback is the computational expense. It achieves a high accuracy
 for the coupling of a lattice Boltzmann model with a
 diffusion-reaction PDE \cite{leemput}.  However, lifting in the
 CR-algorithm requires many additional LBM steps.  This computational
 cost is too expensive to be useful in more complex problems in higher
 dimensions. These drawbacks became clear in \cite{vanderhoydonc} when
 comparing the different methods numerically. The intention of this
 paper is to alleviate them.

In this paper we propose a numerical Chapman-Enskog expansion that
seriously reduces the computational cost of the lifting.  It combines
the idea of the Constrained Runs algorithm with the Chapman-Enskog
expansion and does not need an analytical derivation as the
Chapman-Enskog expansion.  This lifting operator is calculated before
the simulation and finds the coefficients of the Chapman-Enskog
expansion numerically. Once these coefficients are found the
application of the lifting operator is just a stencil computation as
cheap as the analytical Chapman-Enskog expansion.  As a spin-off it
also extracts the macroscopic PDE from the lattice Boltzmann model.
This allows us to construct the hybrid model without deriving the
macroscopic PDE analytically.  The numerical results show that the new
lifting operator can also reach a high accuracy.  Although, we
illustrate and benchmark the new method on academic model problems, we
believe that it is applicable to other discretizations of the Boltzmann
equation.  Furthermore, for the clarity of the presentation we have
kept the boundary between the LBM and PDE domain fixed. In a real
application this boundary might be moved adaptively, triggered by an
error estimate similar as in adaptive mesh refinement. 

This work is organized as follows. In Section
\ref{model} the model problem is defined. It focuses, in particular, on
a hybrid model that consists of a LBM in one part of the domain and
a macroscopic equivalent PDE in another part. Section
\ref{review} gives an overview of existing lifting techniques that are used
in the literature. 
The Chapman-Enskog expansion and the Constrained Runs algorithm are respectively
considered in Sections \ref{chapman_lifting} and \ref{constrained_runs}.
These methods
are discussed in Section \ref{drawbacks}. We tend to remove these
drawbacks by considering a numerical Chapman-Enskog expansion in
Section \ref{chapman_enskog}. 
Section \ref{numerical_results} contains the numerical results. 
In Section \ref{test_problem} the proposed lifting operator is tested in a
setting of restriction and lifting. The application of the lifting
operator to the hybrid LBM and PDE model is considered in Sections
\ref{hybrid_1D} and \ref{hybrid_2D}. We conclude and give an outlook in Section
\ref{conclusion}.

%
%

\section{Model problem} \label{model}
Kinetic models make use of the Boltzmann equation \cite{succi} that describes the evolution of a distribution function $f(\bfx,\bfv,t)$ (function space $C^2_{\mathbb{R}}(D)$) that counts the number of particles or individuals in point $\bfx \in D_{\bfx}\subset \mathbb{R}^n$, $n \in \mathbb{N}_0$, with a velocity $\bfv \in D_{\bfv} \subset \mathbb{R}^n$, at time $t \geq 0$. The equation is
\begin{equation}
\frac{\partial}{\partial t}f(\bfx,\bfv,t)+ \bfv \frac{\partial }{\partial \bfx}f(\bfx,\bfv,t)+F(\bfx,t)\frac{\partial}{\partial \bfv}f(\bfx,\bfv,t)=\Omega.
\label{boltzmann}
\end{equation}
This is an evolution law in phase space where $F(\bfx,t)$ is the external force and $\Omega$ an integral operator that models the reorganization of the velocity distribution due to collisions or other interactions.

The collision operator can be approximated by a simpler Bhatnagar-Gross-Krook (BGK) model 
$\Omega=\omega (f^{eq}(\bfx,\bfv,t)-f(\bfx,\bfv,t))$ \cite{BGK} in which the equilibrium distribution $f^{eq}(\bfx,\bfv,t)$ is given by the Maxwell-Boltzmann distribution \cite{succi}.
The BGK approximation represents a relaxation towards equilibrium with an
associated time scale $\tau=1/\omega$. 

At the moment it is still computationally expensive to simulate or
analyze a Boltzmann model numerically. The development of efficient
numerical methods for models based on the Boltzmann equation is
therefore an active research field. A possible increase in efficiency
can be obtained by constructing a hybrid model. 
The kinetic model is then replaced with a  macroscopic
description in the regions of the spatial domain where this is
justified, for example, away from reaction fronts.  These macroscopic
models are cheaper to simulate.


Possible macroscopic models for fluid dynamics are described by the
Navier-Stokes and Euler equations. One can derive both the
Navier-Stokes as the Euler equations from the Boltzmann equation
\cite{LGCA_LBM}. A lifting operator then transforms the variables
  of the PDE to the distribution function of the Boltzmann equation at
  the boundaries between the domains.

In this paper we study a simple model that allows a detailed
  study of the lifting operator both for the initialization of the distribution function and in a
  hybrid context.  The model uses a lattice Boltzmann model with an
  equilibrium distribution function that only depends on the
  density. While the macroscopic PDE is an advection-diffusion-reaction
  equation for the density only. This simple model allows a detailed
  analysis yet it is general enough to expect that the results can be
  extended to  more realistic Boltzmann models.  The correspondence
  of the lattice Boltzmann with the Boltzmann equation is discussed in
  \cite{LGCA_LBM, sterling_chen}.


\subsection{Lattice Boltzmann models} \label{section_LBM}

A lattice Boltzmann model (LBM) \cite{succi, LGCA_LBM} is a special
discretization of Eq.~\eqref{boltzmann}.  It describes the evolution of
one-particle distribution functions $f_i(\bfx,t)=f(\bfx,\bfv_i,t)$
discretized in space $\bfx$, time $t$ and velocity $\bfv_i$. The
velocities are taken from a discrete set defined by the geometry of
the grid. The functions are represented as $f_i: \mathcal{X} \times
\mathcal{T} \rightarrow \mathbb{R}$ with $\mathcal{X} \times
\mathcal{T} $ the space-time grid with space steps $\Delta \bfx_i$ in
the direction of velocity $\bfv_i$, time step $\Delta t$ and
$\mathcal{T}=\{0,\Delta t, 2\Delta t,\ldots\} \cap [0,T]$.  Representation DdQq used for the description of LBMs stands for d dimensions and q velocity
directions.  D1Q3, for example, considers in a one-dimensional spatial
domain only three values for the velocity $v_i=c_i \Delta x/\Delta t$
with $c_i=i$, $i \in \{-1,0,1\}$ the dimensionless grid velocities.

The remaining of this section contains the description of the lattice
Boltzmann equation in one dimension but can easily be generalized to
more dimensions.

The lattice Boltzmann equation (LBE) describing the evolution of the
distribution functions (
with BGK approximation and no external force in
Eq.~\eqref{boltzmann}) is
\begin{equation}\label{LBE}
f_i(x+c_i\Delta x,t+\Delta t)=(1-\omega)f_i(x,t)+\omega f_i^{eq}(x,t).
\end{equation}
The equilibrium distributions are given by
$f_i^{eq}(x,t)=\frac{1}{3}\rho(x,t)$, $i \in \{-1,0,1\}$
\cite{vandersman} in which the particle density $\rho(x,t)$ is defined
as the zeroth order moment of the distribution functions
$\rho(x,t)=\sum_{i \in \{-1,0,1\}} f_i(x,t)$. These equilibrium
distributions correspond to a local diffusive equilibrium. 

The focus of this paper concerns the initialization of a lattice
Boltzmann model.  Starting the LBM scheme from a given initial density
includes some arbitrariness. The distribution functions $f_i(x,0)$ at time $t=0$
need to be constructed from a given density
$\rho(x,0)$. When the initialization is not consistent it leads to
solutions with steep initial layers \cite{mei}.



\subsection{Macroscopic models for LBMs}\label{macro_LBM}

This section contains descriptions that represent macroscopic equivalent PDEs specific for LBMs.

Partial differential equations model, at a macroscopic scale, the evolution of the moments of the particle distribution functions like density $\rho(x,t)=\sum_i f_i(x,t)$, momentum $\phi(x,t)=\sum_i v_if_i(x,t)$ or energy $\xi(x,t)=\frac{1}{2}\sum_i v_i^2 f_i(x,t)$. 

The transition between the distribution functions and the moments is straightforward since the matrix $M$ below is invertible.
\begin{equation}\label{omzet}
\left(\begin{array}{c}
\rho \\
\phi \\
\xi
\end{array}\right)
=
\left(\begin{array}{c c c}
1 & 1 & 1 \\
1 & 0 & -1 \\
\frac{1}{2} & 0 & \frac{1}{2}
\end{array}\right)
\left(\begin{array}{c}
f_{1} \\
f_0 \\
f_{-1}
\end{array}\right)
=M
\left(\begin{array}{c}
f_{1} \\
f_0 \\
f_{-1}
\end{array}\right).
\end{equation}
When we look at these functions in a point $x$ at time $t$, they can
be represented either as $(f_{1}, f_0, f_{-1})^T \in \mathbb{R}^3$ or
as $(\rho, \phi, \xi)^T \in \mathbb{R}^3$. If we focus on the complete
discretization in space, with $n$ spatial grid points, the function
spaces are $\mathbb{R}^{3\times n}$.

It can be shown that the diffusion PDE and the LBM are macroscopic equivalent \cite{leemput_phd} when considering D1Q3 and
\begin{equation}\label{link_PDE_LBM}
\frac{\partial \rho}{\partial t}=D\frac{\partial^2\rho}{\partial x^2}, \quad D=\frac{2-\omega}{3\omega}\frac{\Delta x^2}{\Delta t}, \quad f_i^{eq}(x,t)=\frac{1}{3}\rho(x,t), \quad i \in \{-1,0,1\}.
\end{equation}
This can be checked by using a Chapman-Enskog expansion. Here $f_i(x,t)$, $i \in \{-1,0,1\}$ is written as a series, each term containing higher order derivatives of $\rho(x,t)$ \cite{chapman_cowling, leemput_phd, cercignani}.
\begin{equation}\label{analytical_chapman}
f_i=f_i^{[0]}+f_i^{[1]}\Delta x+f_i^{[2]} \Delta x^2 + f_i^{[3]}\Delta x^3 +\ldots,
\end{equation}
where
\[
f_i^{[0]}=f_i^{eq}=\frac{\rho}{3}, \quad f_i^{[1]}=-\frac{i}{3\omega}\frac{\partial \rho}{\partial x}, \quad f_i^{[2]}= -\frac{1}{18\omega^2}(\omega-2)(3i^2-2)\frac{\partial^2\rho}{\partial x^2}.
\]
The macroscopic diffusion PDE \eqref{link_PDE_LBM} is obtained by summing the series of the Chapman-Enskog expansion in \eqref{analytical_chapman} over the velocities. This considers purely diffusive effects.

For advection-diffusion problems with uniform velocity field $a$ on D1Q3 the equilibrium distribution functions are given by \cite{vandersman}
\begin{equation}\label{eq:advection}
f_i^{eq}(x,t)=\frac{1}{3}\biggl(1+\frac{i c a}{c_s^2}+\frac{a^2}{c_s^2}\bigg), \quad i\in \{-1,0,1\}, \quad c_s^2=\frac{2}{3}c^2, \quad c=\frac{\Delta x}{\Delta t}, 
\end{equation}
with the equivalent macroscopic description
\begin{equation}\label{link_advection_LBM}
\frac{\partial \rho}{\partial t}+a\frac{\partial \rho}{\partial x}=D\frac{\partial^2\rho}{\partial x^2}, \quad D=\frac{2-\omega}{3\omega}\frac{\Delta x^2}{\Delta t}.
\end{equation}

Similar results can be obtained for higher dimensional problems.
In two spatial dimensions, represented by $x$ and $y$, with equal space steps $\Delta x = \Delta y$, the macroscopic equivalence is given by
\begin{eqnarray} \label{PDE_D2Q5}
& & \frac{\partial \rho}{\partial t}=D\frac{\partial^2\rho}{\partial x^2}+D\frac{\partial^2\rho}{\partial y^2}, \quad D=\frac{2-\omega}{3\omega}\frac{\Delta x^2}{\Delta t}, \quad f_i^{eq}(\bfx,t)=w_i \rho(\bfx,t), \quad i \in \{0,\ldots,4\}, \nonumber \\
& & \bfx = \begin{pmatrix} x \\ y \end{pmatrix} , \quad w_0 = \frac{1}{3}, \quad w_1=\ldots=w_4=\frac{1}{6},
\end{eqnarray}
for D2Q5 \cite{vandersman} and
\begin{eqnarray} \label{PDE_D2Q9}
& & \frac{\partial \rho}{\partial t}+a_x \frac{\partial \rho}{\partial x} +a_y \frac{\partial \rho}{\partial y}=D\frac{\partial^2\rho}{\partial x^2}+D\frac{\partial^2\rho}{\partial y^2}, \quad D=\frac{2-\omega}{3\omega}\frac{\Delta x^2}{\Delta t}, \nonumber \\
& & f_i^{eq}(\bfx ,t)=w_i \rho \biggl(1+\frac{\bfv_i \cdot \mathbf{a} }{c_s^2}+\frac{(\bfv_i \cdot \mathbf{a} )^2}{2c_s^4}-\frac{\mathbf{a}^2}{2c_s^2}\bigg), \quad i \in \{0,\ldots,8\}, \quad \mathbf{a} =(a_x,a_y), \quad c_s^2=\frac{1}{3}\frac{\Delta x^2}{\Delta t^2}, \nonumber \\
& & \bfx = \begin{pmatrix} x \\ y \end{pmatrix}, \quad w_0=\frac{4}{9}, \quad w_i=\frac{1}{9}, \quad i\in \{1,\ldots,4\}, \quad w_i=\frac{1}{36}, \quad i \in \{5,\ldots,8\},
\end{eqnarray}
for D2Q9 \cite{vandersman}.

With such analytical expressions from the Chapman-Enskog expansion available, a lifting operator can be constructed. Indeed, $f_i(x,t)$ is then written as a series in function of the given density $\rho(x,t)$. This allows us to construct the distribution functions from a given initial density necessary to initialize the LBM.




We will consider the lattice Boltzmann model as the `exact' model and
these PDEs of the density as the macroscopic approximations. From this, we can
construct a hybrid model problem as outlined in Section \ref{hybrid_models}.

\subsection{Hybrid models} \label{hybrid_models} 
\begin{figure}[!htop]
\begin{center}
\includegraphics[width=0.7\textwidth]{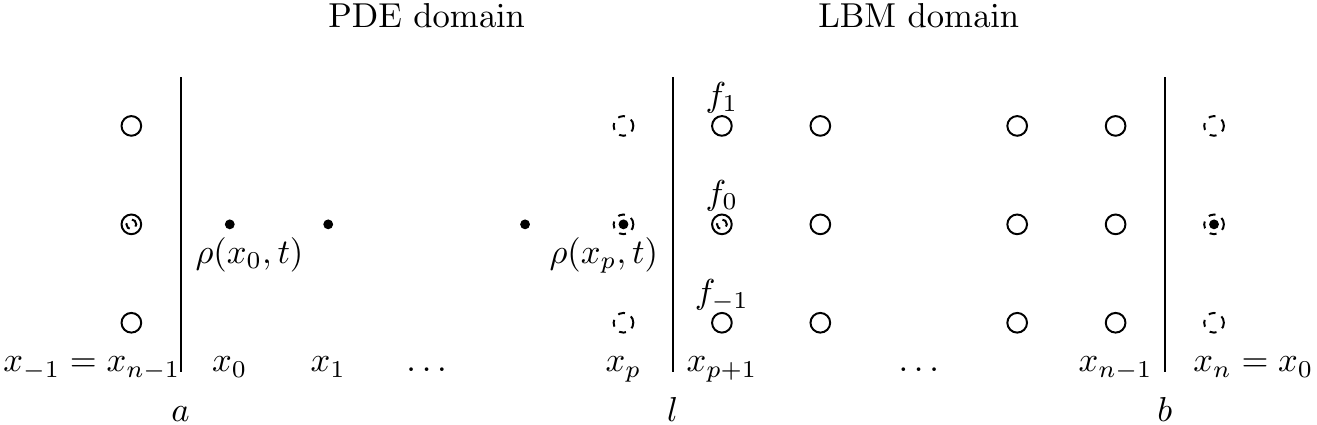}
\end{center}
\caption{ The domain $[a,b]$ in the hybrid model is split into $[a,l[$
  on which we solve the PDE model and $[l,b]$ on which we solve the
  LBM.  The solid points ($\bullet$) represent the grid for the
  density $\rho$ of the discrete PDE, the circles ($\circ$) represent
  the LBM variables $(f_{1},f_0,f_{-1})^T$. The periodic boundary
  conditions and the coupling are implemented with ghostcells which
  are drawn by dashed circles. The density in the ghostcells of the
  PDE domain, in $x_{-1}$ and $x_{p+1}$, are found by taking $\sum_i
  f_i$ in $x_{n-1}$ and $x_{p+1}$, respectively.  However, the
  ghostcells for the LBM domain, in $x_p$ and $x_{n}$, require a
  lifting operator that lifts $\rho$ to $(f_{1},f_0,f_{-1})^T$ in
  these points. \label{path}}
\end{figure}

This section deals with the construction of the hybrid model problem.
Consider the one-dimensional problem D1Q3 but bear in mind that a
similar construction of a hybrid model can be done in higher
dimensions.  In particular, we couple a lattice Boltzmann model with
the macroscopic equivalent PDE. The resulting
hybrid domain for D1Q3 is presented in Figure \ref{path}.  A
one-dimensional domain $[a,b]$ is considered that couples the PDE \eqref{link_PDE_LBM} on
$[a,l[$ with the LBE \eqref{LBE} on $[l,b]$. Furthermore, we assume periodic boundary conditions. Note that in both
subdomains the same grid spacings are used in space ($\Delta x$) and
in time ($\Delta t$) and the boundary $l$ remains fixed for all
times. Using a different space-time grid is of particular interest for
future work but the same spacings are used to highlight the coupling
error. Similarly, the boundary may be moved adaptively as in adaptive
mesh refinement. In an actual physical problem it will be important to
use the hybrid model that gives a Boltzmann description where shock
waves, contact discontinuities or sharp gradients occur. Since these
move in time it might be useful to work with a moving interface method
as in \cite{moving_interface} and \cite{multiscale_degond}. However,
this is not the focus of the current paper.

On the domain $[a,l[$, we discretize the PDE \eqref{link_PDE_LBM} with cell centered
central differences in space and forward Euler time discretization. The grid
points $x_j$ with $j \in \{0,1,\ldots,p\}$ cover this domain and for
these points it holds that
\begin{eqnarray}
\rho(x_j, t + \Delta t) = \rho(x_j,t)+ \frac{D\Delta t}{\Delta x^2} \left(\rho(x_{j-1},t) -2\rho(x_j,t) + \rho(x_{j+1},t)\right).
\end{eqnarray}
For the grid points $x_j$ with $j \in \{p+1,\ldots, n-1\}$ in $[l,b]$
the LBE \eqref{LBE} holds.

The
full domain has an initial condition $\rho(x_j,0)$, $\forall j \in
\{0,1,\ldots,n-1\}$. The lifting operator is required to formulate the
initial conditions of the LBM domain $f_i(x_j,0)$, $\forall j \in
\{p+1,\ldots,n-1\}$.

The periodic boundary conditions lead to the following boundary
conditions for the PDE domain: $\forall t : \rho(x_{-1},t) = \sum_i
f_i(x_{n-1},t)$ and $\rho(x_{p+1},t) = \sum_i f_i(x_{p+1},t)$.

The aim is to construct the boundary conditions of the LBM domain in such a way
that $\forall t > 0$ and $\forall j \in \{0,1,\ldots,n-1\}$ the macroscopic density defined as
\begin{equation}
  \rho(x_j,t) = 
\begin{cases}
  \rho(x_j,t)  \quad&\text{if}\quad j \in \{0,\ldots,p\},\\
\sum_i f_i(x_j,t) \quad&\text{if}\quad  j \in \{p+1,\ldots,n-1\},\\
\end{cases}
\end{equation}
behaves as the density of a LBM solved on the full domain. 

To formulate these boundary conditions, a lifting operator is required
that maps the density $\rho(x,t)$ in the ghost points $x_0$ and $x_p$,
the unknown of the PDE, to the distribution functions $f_i(x,t)$, $i
\in \{-1,0,1\}$ of the LBM.

This can be generalized by considering a higher dimensional spatial domain. 
The remaining derivations in this paper focus on one dimension although they can also be generalized to more dimensions.

%
%

\section{Review of existing lifting operators} \label{review} This
section gives an overview of existing lifting operators that map densities to
distribution functions. An analytical expansion that
expresses the distribution functions as a series of the density and
its spatial derivatives is given in Section
\ref{chapman_lifting}, while a numerical method is presented in
Section \ref{constrained_runs}. Section \ref{drawbacks} 
discusses these methods which results in the motivation to 
propose a lifting operator based on the combined ideas of the
analytical and numerical method.


\subsection{Chapman-Enskog expansion} \label{chapman_lifting} The
Chapman-Enskog expansion \cite{chapman_cowling, leemput_phd,
  cercignani}, already discussed in Section \ref{hybrid_models}, can
be used as a lifting operator.  It constructs a mapping from
$\rho(x,t)$ to $f_i(x,t)$, $i \in \{-1,0,1\}$ (D1Q3). 



\subsection{Constrained Runs algorithm} \label{constrained_runs} 
An alternative numerical procedure is the Constrained Runs algorithm discussed in this section.
It is
well known that in phase space the dynamics are quickly attracted
toward a slow manifold \cite{gear}. For the problem we are studying
the dynamics on the slow manifold can be parameterized by the density
$\rho(x,t)$.  The distribution functions are then of the form
$\{f_1(\rho(x,t)),f_0(\rho(x,t)),f_{-1}(\rho(x,t))\}$.

Because of Eq.~\eqref{omzet}, it is equivalent to determine
$\{f_{1}(\rho),f_0(\rho),f_{-1}(\rho)\}$ or
$\{\rho,\phi(\rho),\xi(\rho)\}$. The missing
distribution functions $\{f_{1},f_0,f_{-1}\}$ can be found by determining
$\phi$ and $\xi$ for a given $\rho$ such that $\phi$, $\xi$ and $\rho$
lie on the slow manifold. This is the basic idea of the Constrained
Runs (CR) algorithm that was proposed by Gear \textit{et al.}~\cite{gear} 
for stiff singularly perturbed ordinary differential
equations (ODEs) to map  macroscopic initial data to missing
microscopic variables.  It uses the numerical simulator to find the
missing data such that the evolution is close to the slow manifold.

This Constrained Runs algorithm can be applied to lattice Boltzmann
models \cite{initialization}. The state of the lattice Boltzmann model
can be split into
\begin{equation*}
  u = \left( \rho \right)   \quad \text{and} \quad  v=\left(\begin{array}{c}\phi \\ \xi \end{array}\right),
\end{equation*}
where $u \in \mathbb{R}^n$ and $v \in \mathbb{R}^{2n}$ for a LBM with
$n$ spatial grid points. The density $\rho$ is known so $u$ is
given, while $v$ is unknown since $\phi$ and $\xi$ are missing. Denote the known 
initial conditions as $u(0)=u_0 \in \mathbb{R}^n$.

\begin{figure}[!htop]
\begin{center}
\includegraphics[width=0.6\textwidth]{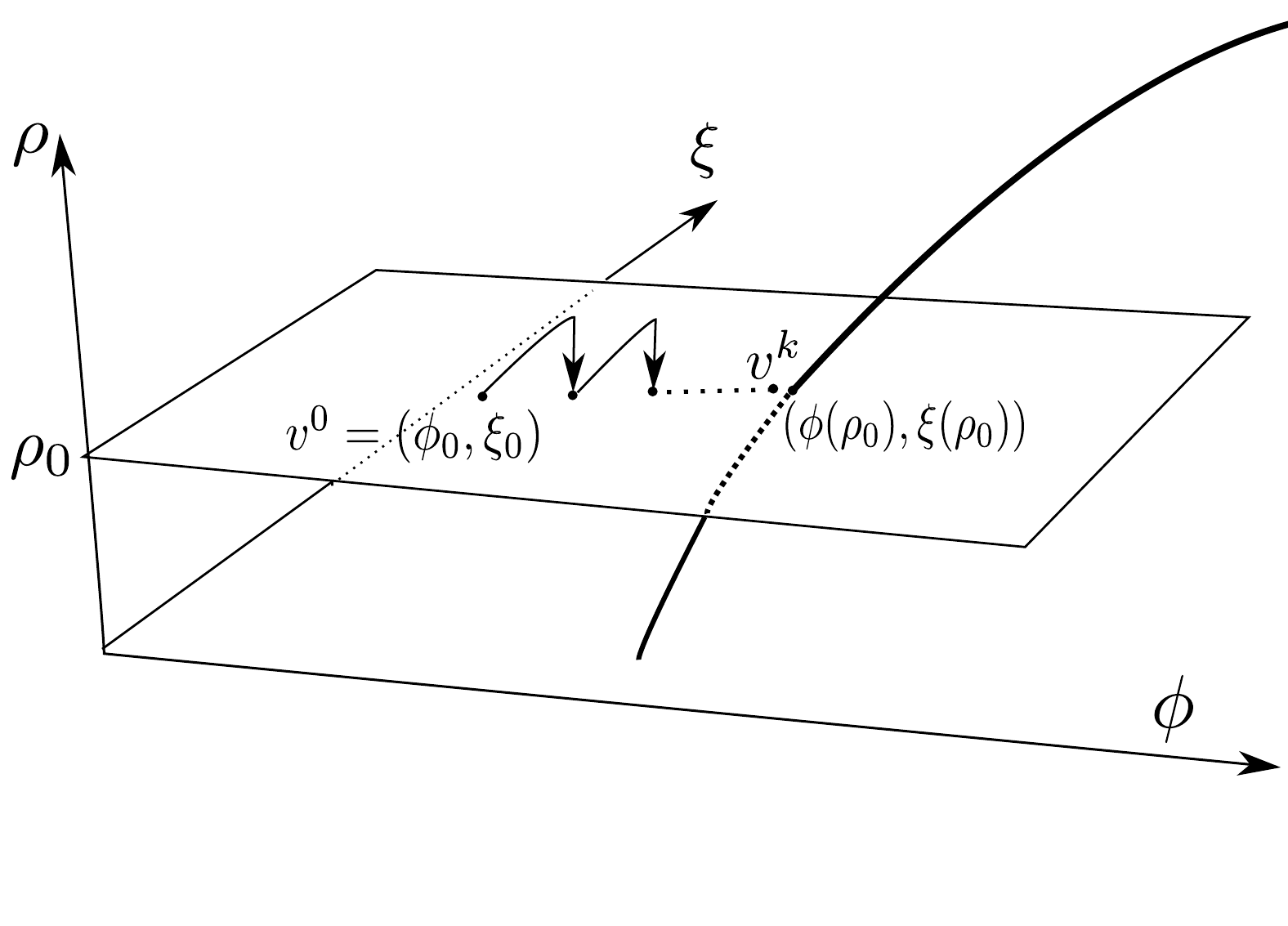}
\end{center}
\caption{ Sketch of the first few steps of the Constrained Runs
  algorithm for the LBM with a constant backward extrapolation in
  time.  The solid line shows the evolution of
  $\{\rho(x,t),\phi(x,t),\xi(x,t)\}$ along the slow manifold for a
  given grid point $x$.  For a given $\rho_0$ we search for the
  intersection of the plane with the slow manifold.  We start
  iterating with $\rho_0$, the known density, and initial guesses
  $\phi_0$ and $\xi_0$ for the missing moments ($v^0$). After each
  step of the LBM, the density is reset to its initial value $\rho_0$
  but the moments evolve during the LBM time simulation. This results
  in $v^k$, the $k$-th iterate of the
  CR-algorithm. This algorithm finds an approximation for the missing
  values $\phi$ and $\xi$ on the slow manifold. \label{CR}}
\end{figure}

The idea is now to initialize $v$ such that the evolution of $v$ under the LBM is
smooth of order $m$. The smoothness condition is defined by
\begin{equation*}
\frac{d^{m+1}v(t)}{dt^{m+1}}\bigg|_{t=0}=0,
\end{equation*}
which is approximated  by
\begin{equation}
\Delta^{m+1}v(t) \approx \Delta t^{m+1} \frac{\mathrm{d}^{m+1}v(t)}{\mathrm{d}t^{m+1}},
\end{equation}
where $\Delta^m$ is the well-known forward finite difference stencil
on $v(t)$, $v(t + \Delta t)$, $\ldots$.  For $m=0$, the converged $v$
satisfies the smoothness condition, up to a certain tolerance, and it
is an approximation to the point of intersection with the slow
manifold. This is schematically represented in Figure \ref{CR}.  This
iteration is always stable and the point of intersection is found to
first order accuracy compared to the Chapman-Enskog expansion for the LBM
with BGK collisions for one-dimensional reaction-diffusion problems
\cite{initialization}.  For $m \ge 1$, multiple LBM steps are
necessary to estimate the derivative. For $m=1$, this is often
interpreted as a backward linear extrapolation in time
\cite{vandekerckhove}. 

The number of LBM steps used in the backward extrapolation determines
the accuracy of the scheme. Higher order schemes increase the accuracy
but they can become unstable.  In \cite{vandekerckhove} this
instability is circumvented by formulating the point of intersection
as a fixed point 
\begin{equation} \label{eq:CR}
v^{k+1}=\mathcal{C}_m(u_0,v^k),
\end{equation}
where $\mathcal{C}_m$ denotes one step of the CR-algorithm and $m$ is
related to the order of the time derivative that is set to zero in the
backward extrapolation in time. In general Eq.~\eqref{eq:CR} is
nonlinear and the fixed point can be found by a Newton-Krylov
iteration. However, this requires many additional LBM evaluations to
construct the Jacobian. Similarly, matrix-free methods like
GMRES still require many matrix-vector products since the spectrum is
unfavorable for fast convergence \cite{vandekerckhove_phd}. In \cite{vanderhoydonc} the CR-algorithm is
combined with Newton's method by performing local updates at the ghost
points of the hybrid model to reduce the size of the Jacobian.

\subsection{Discussion of existing lifting
  operators} \label{drawbacks} The methods discussed in Sections
\ref{chapman_lifting} and \ref{constrained_runs} are well known
methods to construct a lifting operator for LBMs. However, each of these
methods has some drawbacks.  As noted earlier, a drawback of the use
of the Chapman-Enskog expansion (Section \ref{chapman_lifting}) is the
necessity to construct the expressions analytically. Therefore its use
is limited to a few examples where the expansion is known.
However, its computational cost is limited to the calculation of
the numerical approximation of the derivatives.  The Chapman-Enskog
expansion then becomes a stencil operator and the cost of the
application grows linearly with the number of points where lifting is
required.

The Constrained Runs scheme (Section \ref{constrained_runs}) can be
used to approximate these expressions numerically.  However, the
lifting method can become computationally expensive since it requires
many evaluations of the underlying lattice Boltzmann model to
construct the Jacobian matrix. Even with the matrix-free methods and local updates discussed at the end of Section \ref{constrained_runs} it still remains computationally
expensive to use in practice, especially in higher dimensional
problems. 



As an advantage of the Constrained Runs algorithm, we should note that
the lifting error can be smaller than the modeling error, the difference
in density between the LBM and its PDE approximation, by using the
CR-algorithm in the hybrid model discussed in Section
\ref{hybrid_models} \cite{leemput}.  Section \ref{section_compare}
contains a comparison of the computational cost of these existing
methods in the sense of hybrid models.

The focus of this paper is to obtain an alternative lifting operator
that reduces the computational cost but holds the advantage of
achieving the modeling error.

%
%

\section{Numerical Chapman-Enskog expansion} \label{chapman_enskog} In
this section, we construct a lifting operator that alleviates the
computational expense of the CR-algorithm. It combines the ideas of
Constrained Runs and the Chapman-Enskog expansion. Instead of using
Constrained Runs to find for each grid point the missing moments
$\phi$ and $\xi$ of the distribution functions, we use Constrained
Runs to find the unknown coefficients of the Chapman-Enskog
expansion. This has several advantages that will be discussed at the
end of Section \ref{algorithm}.

The derivations in this section are again based on one-dimensional problems but can easily be generalized to more dimensions.

\subsection{Distribution functions as a series of the density}
This section shows that the solution $f_i(x,t)$, $i\in
\{-1,0,1\}$ of a LBM with an infinite domain and parameters $\Delta
x$, $\Delta t$ and $\omega$ can be written as a series of $\rho(x,t)$,
the macroscopic density. We initially characterize distribution
functions $f_i$ as smooth functions that are sufficiently
differentiable functions in time and space which implies that the same
holds for the density, a sum of these distribution functions.  The
smoothness condition will be specified below. This condition can be justified
when the lattice spacing $\Delta x$ is much bigger than the mean free
path \cite{phd_amsterdam, cellular_automata}.   Then
the distribution functions can be written as
\begin{eqnarray} \label{constants}
  f_i(x,t) & = & f_i^{eq}(x,t) + \alpha_i \frac{\partial \rho}{\partial x} + \beta_i  \frac{\partial^2 \rho}{\partial x^2} + \delta_i \frac{\partial^3 \rho}{\partial x^3}+ \epsilon_i \frac{\partial^4 \rho}{\partial x^4} + \ldots \nonumber \\
  & & + \gamma_i \frac{\partial \rho}{\partial t} + \zeta_i \frac{\partial^2 \rho}{\partial t^2} + \ldots + \eta_i \frac{\partial^2 \rho}{\partial x \partial t} + \ldots,
\end{eqnarray}
where \begin{equation} \label{vectors_constants}
\alpha=
\left(\begin{array}{c}
\alpha_1 \\
\alpha_0 \\
\alpha_{-1}
\end{array}\right) \in \mathbb{R}^3, \quad
\beta=
\left(\begin{array}{c}
\beta_1 \\
\beta_0 \\
\beta_{-1}
\end{array}\right) \in \mathbb{R}^3,
 \ldots,
\end{equation}
are fixed constants that only depend on $\omega$, $\Delta x$ and
$\Delta t$. The derivation of this expansion is outlined in the remaining of this section.

Since the functions $f_i(x,t)$ are infinitely differentiable, a Taylor
expansion can be constructed. The distribution functions in point
$x+i\Delta x$, $i \in \{-1,0,1\}$ at time $t+\Delta t$ are given by
\[
f_i(x+i\Delta x, t+\Delta t)=f_i(x,t)+\frac{\partial f_i}{\partial x}
i\Delta x+\frac{\partial^2 f_i}{\partial x^2} \frac{i^2 \Delta x^2}{2}
+ \frac{\partial f_i}{\partial t} \Delta t + \frac{\partial^2
  f_i}{\partial t^2} \frac{\Delta t^2}{2} + \frac{\partial^2
  f_i}{\partial x\partial t} i\Delta x \Delta t + \ldots. \] Combined
with the assumption that $f_i$ is a solution of the LBE
\eqref{LBE} on an infinite domain, we end up with
\[
f_i(x,t)=f_i^{eq}(x,t)-\frac{i\Delta x}{\omega}\frac{\partial f_i}{\partial x}-\frac{i^2 \Delta x^2}{2\omega}\frac{\partial^2 f_i}{\partial x^2}-\frac{\Delta t}{\omega}\frac{\partial f_i}{\partial t}-\frac{\Delta t^2}{2\omega}\frac{\partial^2 f_i}{\partial t^2} -\frac{i\Delta x \Delta t}{\omega} \frac{\partial^2 f_i}{\partial x \partial t}-\ldots.
\]
With the notation $\mathcal{L}_i$ for the functional
\begin{equation} \label{functional}
\mathcal{L}_i=-\frac{i\Delta x}{\omega}\frac{\partial }{\partial x}-\frac{i^2 \Delta x^2}{2\omega}\frac{\partial^2 }{\partial x^2}-\frac{\Delta t}{\omega}\frac{\partial }{\partial t}-\frac{\Delta t^2}{2\omega}\frac{\partial^2 }{\partial t^2} -\frac{i\Delta x \Delta t}{\omega} \frac{\partial^2 }{\partial x \partial t}-\ldots,
\end{equation}
we can rewrite the LBE into a set of three coupled PDEs for the distribution functions
\begin{equation}
(1-\mathcal{L}_i) f_i(x,t)=f_i^{eq}(x,t), \quad  \forall i \in \{-1,0,1\},
\end{equation} 
that holds for $x \in
]-\infty,\infty[$ and $t \in [0,\infty[$.

The solution can be found
by performing a Picard or fixed point iteration 
\begin{equation}\label{eq:picarditeration}
f_i^{(n+1)} = \mathcal{L}_if_i^{(n)} + f_i^{eq},
\end{equation}
with initial guess $f_i^{(-1)} = 0$ that results in
\begin{eqnarray*}
f_i^{(0)} &=& f_i^{eq} =: g_i,\\
f_i^{(1)} &=& g_i+\mathcal{L}_i f_i^{(0)} = g_i+\mathcal{L}_ig_i,\\
f_i^{(2)} &=& g_i+\mathcal{L}_ig_i+\mathcal{L}_i(\mathcal{L}_ig_i),\\
&\ldots& \\
f_i^{(n)} &=& \sum_{k=0}^n \mathcal{L}_i^k g_i,
\end{eqnarray*}
with $f_i^{(n)}$ the $n$-th iterate.  This iteration converges if the
error between subsequent iterations goes to zero. 

In contrast to traditional iterations, which require convergence for any initial guess,
Eq.~\eqref{eq:picarditeration} is a fixed point iteration with
initial guess zero and a smooth right hand side. It is only necessary
to show convergence for this particular case. To discuss this convergence
we introduce the 2-norm, $\|.\|$, to show what happens between subsequent iterations.
The absolute difference of subsequent iterations is given by
\[
\|f_i^{(n+1)}-f_i^{(n)}\|=\|\mathcal{L}_i^{n+1} g_i\|.
\]
This goes to zero if $f_i$ is smooth enough, implying smoothness on
$\rho$ and $g_i=f_i^{eq}(x,t)$ such that $\lim_{n\rightarrow \infty}
\|\mathcal{L}_i^{n+1} g_i\| =0$. This smoothness condition depends on
the parameters of the LBM, $\Delta x$, $\Delta t$ and $\omega$,
and the derivatives of $g_i$.  For example, when $g_i$ can be
described by a polynomial, we have that there exists a $k$ such that
for all $n > k$ applies that $\|\mathcal{L}_i^{n+1} g_i\|=0$.

We end up with the series \eqref{constants} that consists of the
vectors of constants given in \eqref{vectors_constants}, the density
and its derivatives.
Once the constants are determined, the lifting operator --- that is necessary to initialize the LBM and to determine the ghost points in the hybrid model --- can be constructed. How these constants are found is discussed in Sections \ref{det_constants} and \ref{spatial_der}. Section \ref{det_constants} deals with the analytical derivation while Section \ref{spatial_der} is concerned with the numerical procedure. As a surplus, it allows us to find the corresponding macroscopic PDE as outlined in Section \ref{der:PDE}.
\subsection{Derivation of a lifting operator} 
\label{det_constants}
With the help of expansion \eqref{constants} it is possible
  to build  a lifting operator that constructs the distribution
  functions for a given density. The focus of this section
lies in the determination of the vectors of constants
\eqref{vectors_constants}, the coefficients of such a lifting operator
\eqref{constants}. To simplify the discussion and notation we limit ourselves
  to a truncated series
\begin{equation} \label{eq_coeff}
f(x,t)=f^{eq}(x,t)+\alpha \frac{\partial \rho}{\partial x} + \beta  \frac{\partial^2 \rho}{\partial x^2}+\gamma \frac{\partial \rho}{\partial t},
\end{equation}
where $\alpha$, $\beta$ and $\gamma$ are the vectors containing the
  constants.  The method is easily generalized to include higher order
  terms which will be considered in Section \ref{sec:higherorder}.

Using the fact that Eq.~\eqref{eq_coeff} is valid for every
possible grid point, we can consider three grid points $x_j$, $x_k$
and $x_l$ and set up a linear system for the nine unknowns, namely three vectors each containing three constants. Where $j$, $k$ and
$l$ are certain indices determined in a later stage of the paper.  {\fontsize{4.99999}{7}\selectfont
\begin{eqnarray} \label{system_constants}
\left(\begin{array}{ccc|ccc|ccc}
\frac{\partial \rho(x_j)}{\partial x} & & & \frac{\partial^2 \rho(x_j)}{\partial x^2}&&& \frac{\partial \rho(x_j)}{\partial t} &&\\
&\frac{\partial \rho(x_j)}{\partial x} & & & \frac{\partial^2 \rho(x_j)}{\partial x^2}&&& \frac{\partial \rho(x_j)}{\partial t} &\\
&&\frac{\partial \rho(x_j)}{\partial x} & & & \frac{\partial^2 \rho(x_j)}{\partial x^2}&&& \frac{\partial \rho(x_j)}{\partial t} \\
\hline
\frac{\partial \rho(x_k)}{\partial x} & & & \frac{\partial^2 \rho(x_k)}{\partial x^2}&&& \frac{\partial \rho(x_k)}{\partial t} &&\\
&\frac{\partial \rho(x_k)}{\partial x} & & & \frac{\partial^2 \rho(x_k)}{\partial x^2}&&& \frac{\partial \rho(x_k)}{\partial t} &\\
&&\frac{\partial \rho(x_k)}{\partial x} & & & \frac{\partial^2 \rho(x_k)}{\partial x^2}&&& \frac{\partial \rho(x_k)}{\partial t} \\
\hline
\frac{\partial \rho(x_l)}{\partial x} & & & \frac{\partial^2 \rho(x_l)}{\partial x^2}&&& \frac{\partial \rho(x_l)}{\partial t} &&\\
&\frac{\partial \rho(x_l)}{\partial x} & & & \frac{\partial^2 \rho(x_l)}{\partial x^2}&&& \frac{\partial \rho(x_l)}{\partial t} &\\
&&\frac{\partial \rho(x_l)}{\partial x} & & & \frac{\partial^2 \rho(x_l)}{\partial x^2}&&& \frac{\partial \rho(x_l)}{\partial t}
\end{array}
\right)
\left(\begin{array}{c}
  \alpha_1\\
  \alpha_0\\
  \alpha_{-1}\\\hline
  \beta_{1}\\
  \beta_0\\
  \beta_{-1}\\\hline
  \gamma_{1}\\
  \gamma_0\\
  \gamma_{-1}\\
\end{array}
\right) \nonumber \\
=
\left(\begin{array}{c}
  f_{1}(x_j,t) - f^{eq}_{1}(x_j,t)\\
  f_0(x_j,t) - f^{eq}_0(x_j,t)\\
  f_{-1}(x_j,t) - f^{eq}_{-1}(x_j,t)\\\hline
  f_{1}(x_k,t) - f^{eq}_{1}(x_k,t)\\
  f_0(x_k,t) - f^{eq}_0(x_k,t)\\
  f_{-1}(x_k,t) - f^{eq}_{-1}(x_k,t)\\\hline
  f_{1}(x_l,t) - f^{eq}_{1}(x_l,t)\\
  f_0(x_l,t) - f^{eq}_0(x_l,t)\\
  f_{-1}(x_l,t) - f^{eq}_{-1}(x_l,t)\\
\end{array}
\right).
\end{eqnarray}
}

For a given $f_i(x,t)$ where $i \in \{-1,0,1\}$, the linear
  system \eqref{system_constants} will give the coefficients $\alpha$,
  $\beta$ and $\gamma$.  However, linear system
  \eqref{system_constants} only delivers the correct coefficients if
  $f_i$ is smooth enough such that $f_i^{eq}$ satisfies the smoothness
  condition.  This is the case when $f_i$ lies on the slow manifold.
  The Constrained Runs algorithm offers a way to reach the slow
  manifold in an iterative way.
  
We combine the ideas of the CR-algorithm to reach the slow
  manifold and the Chapman-Enskog expansion to find the unknown
  constants on this slow manifold. The numerical procedure to do so is
  given in Section \ref{spatial_der}.

If a PDE in closed form exists that describes the evolution of
  $\rho$ in the form of $\rho_t + a \rho_x = D \rho_{xx}$, then the
  linear system \eqref{system_constants} will be singular. Indeed, the
  PDE will give a relation between $\rho_t$, $\rho_x$ and $\rho_{xx}$
  in each of the grid points $x_j$, $x_k$ and $x_l$. As a result,
  every element in the last three columns of the linear system
  \eqref{system_constants} can be written as a linear combination of
  the first six columns. In practice, however, the PDE is only an
  approximation and the system will be close to singular.


This clarifies why we do not solve the linear system in \eqref{system_constants} and first focus on one that is not close to singular in Section \ref{spatial_der}. Section \ref{der:PDE} explains why such a PDE exists.

%
%
\subsection{Numerical procedure to construct the lifting operator}
\label{spatial_der}
From the previous discussion it is clear that the coefficients of the
lifting operator can be extracted from a linear system once $f$
approaches the slow manifold.  Next, the extraction of the
coefficients is combined with the CR-algorithm to bring $f$ close to
the slow manifold.

\subsubsection{Coefficients of the Chapman-Enskog expansion as a fixed point}
\label{sec:distr_slow}

This discussion is limited, as in Section \ref{det_constants}, to the
first few terms of the expansion.  The singular system can be avoided
by taking a series that only contains spatial derivatives. Such a
series can represent the same state since the time derivative
$\partial_t \rho$ is often related to the spatial derivatives through
a macroscopic PDE.  For example, suppose that a PDE of the form
$\rho_t+a\rho_x=D\rho_{xx}$ describes the behavior of $\rho$. It is
then possible to eliminate $\partial_t \rho$ from the expansion. The
coefficients are then $\tilde{\alpha} = \alpha - \gamma a$ and
$\tilde{\beta} = \beta + \gamma D$.  The distribution functions are now series with only spatial derivatives. 
\begin{equation}\label{f_constant}
f(x,t)=f^{eq}(x,t)+\tilde{\alpha}\frac{\partial \rho}{\partial x}+\tilde{\beta}\frac{\partial^2 \rho}{\partial x^2}.
\end{equation}
\begin{remark}
Rewrite $\tilde{\alpha}$ and $\tilde{\beta}$ as $\alpha$ and $\beta$ but bear in mind that it considers different coefficients in Eq.~\eqref{eq_coeff} and Eq.~\eqref{f_constant}.
\end{remark}

Again, once the distribution functions are close to the slow manifold,
we can extract the coefficients $\alpha$ and $\beta$ from the linear
system
\begin{eqnarray} \label{system_constants_spatial}
\left(\begin{array}{ccc|ccc}
\frac{\partial \rho(x_j)}{\partial x} & & & \frac{\partial^2 \rho(x_j)}{\partial x^2}&& \\
&\frac{\partial \rho(x_j)}{\partial x} & & & \frac{\partial^2 \rho(x_j)}{\partial x^2}& \\
&&\frac{\partial \rho(x_j)}{\partial x} & & & \frac{\partial^2 \rho(x_j)}{\partial x^2} \\
\hline
\frac{\partial \rho(x_k)}{\partial x} & & & \frac{\partial^2 \rho(x_k)}{\partial x^2}&&\\
&\frac{\partial \rho(x_k)}{\partial x} & & & \frac{\partial^2 \rho(x_k)}{\partial x^2}&\\
&&\frac{\partial \rho(x_k)}{\partial x} & & & \frac{\partial^2 \rho(x_k)}{\partial x^2} \\
\end{array}
\right)
\left(\begin{array}{c}
  \alpha_{1}\\
  \alpha_0\\
  \alpha_{-1}\\\hline
  \beta_{1}\\
 \beta_0\\
  \beta_{-1}\\
\end{array}
\right) \nonumber \\
=
\left(\begin{array}{c}
  f_{1}(x_j,t) - f^{eq}_1(x_j,t)\\
  f_0(x_j,t) - f^{eq}_0(x_j,t)\\
  f_{-1}(x_j,t) - f^{eq}_{-1}(x_j,t)\\\hline
  f_{1}(x_k,t) - f^{eq}_{1}(x_k,t)\\
  f_0(x_k,t) - f^{eq}_0(x_k,t)\\
  f_{-1}(x_k,t) - f^{eq}_{-1}(x_k,t)\\
\end{array}
\right).
\end{eqnarray}

To reach the slow manifold we combine Constrained Runs with the
extraction of the coefficients.  Consider a numerical function
$h(\alpha,\beta; \rho, m)$ as described in Function \ref{fun:h}.  
This function takes as input $\alpha$ and
$\beta$ and as parameters a fixed density $\rho$ and an integer $m$,
the order of the smoothness condition.  It first constructs, with this
input, a state $f$ with the help of series \eqref{f_constant}. This
state is then used to perform multiple LBM steps.  For each of these
steps we can find the corresponding moments $\phi$ and $\xi$.  On the
moments, we can use the CR-algorithm (Section \ref{constrained_runs})
to find new moments that are closer to the slow manifold by
considering the finite difference approximations of the $m$-th order
smoothness condition,
\begin{equation} \label{CR_formulas}
\frac{d^{m+1}{\phi}}{d t^{m+1}}=0 \quad \text{and} \quad \frac{d^{m+1}{\xi}}{d t^{m+1}}=0. 
\end{equation}
These new moments result in new coefficients $\alpha$ and $\beta$, by
applying the linear system Eq.~\eqref{system_constants_spatial} on
the distribution functions $f_i$, corresponding to the new $\phi$ and
$\xi$ and the given $\rho$.

The idea is now to determine $\alpha$ and $\beta$ such that they are
invariant under this numerical function $h(\alpha,\beta ;
\rho,m)$. Indeed, if the initial and final state can be described by
the same $\alpha$ and $\beta$ then the lifted $f$ is close to the slow
manifold since  it is a fixed point of the underlying CR-iteration.

Instead of performing a regular fixed point iteration with $h(\alpha,
\beta ; \rho,m)$, a Newton iteration is used that finds $a:=(\alpha,
\beta) \in \mathbb{R}^6 $ such that $a = h(a;\rho,m)$. This reduces
the computational cost significantly because the size of the Jacobian
system with $\alpha$ and $\beta$ is much smaller then the Jacobian of
the original Constrained Runs algorithm. The latter involves
the moments in every grid point and this becomes very large.
\begin{algorithm}[!ht]
\floatname{algorithm}{Function}
  \caption{$h(\alpha,\beta ; \rho, m)$ \label{fun:h}} 
  \label{algo:coefficients}
  \begin{algorithmic}[1]
  \REQUIRE Guess on coefficients $\alpha$, $\beta$, given density $\rho$, order $m$ to use in Eq.~\eqref{CR_formulas}.
    \STATE Construct lifting operator $f = f^{eq} + \alpha \frac{\partial \rho}{\partial x}
    + \beta\frac{\partial^2 \rho}{\partial x^2}$ (Eq.~\eqref{f_constant}).
    \STATE Compute corresponding moments $\phi$ and $\xi$ by applying Eq.~\eqref{omzet}. 
	\STATE Perform $m+1$ LBM time steps to compute $\frac{d^{m+1}{\phi}}{d t^{m+1}}=0$ and
$\frac{d^{m+1}{\xi}}{d t^{m+1}}=0$ by using forward finite difference formulas. 
This results in new moments $\phi$ and $\xi$. \COMMENT{find moments closer to slow manifold}
\STATE Revert back to distribution functions $f$ by applying Eq.~\eqref{omzet}.
    \STATE Select grid points $x_j$, $x_k$ and $x_l$ to construct linear system \eqref{system_constants_spatial}.
    \STATE Solve the system for $\alpha$ and $\beta$.
    \RETURN $\alpha$, $\beta$.
  \end{algorithmic}
\end{algorithm}

The numerical function $h(\alpha,\beta ; \rho,m)$ has a density
$\rho(x,0)$ as a parameter and the solution for the coefficients
is independent of its choice of $\rho$.  The coefficients are determined by functional $\mathcal{L}_i$ \eqref{functional} that only depends
on constants $\omega$, the spatial grid size $\Delta x$ and time step
$\Delta t$. Since the coefficients do not depend on time, we can
choose an arbitrary density $\rho(x,0)$ such that Eq.~\eqref{system_constants_spatial} is easily solvable.
  
The choice of the grid points $x_j$ and $x_k$ in
\eqref{system_constants_spatial} should be such that the condition
number of the matrix is optimal. In addition, the spatial derivatives
that are considered needs to exist and should not become zero during
the LBM evolution since otherwise we would end up with singular linear
systems.

Furthermore, the test domain used in the LBM inside the function
$h(\alpha, \beta ; \rho,m)$ can be significantly smaller than the
domain of the original LBM problem.  A smaller test domain will not
affect the constant coefficients of the lifting operator.  However, it
should use the same $\Delta x$ and $\Delta t$ as the LBM of interest
since the coefficients depend on the chosen spacings in space and
time. The choice for the test domain, density and indices is further
discussed in Section \ref{test_problem} for the considered model problem.

%
%
\subsubsection{Higher order versions}\label{sec:higherorder}
There are two ways to increase the accuracy. First, more terms in the
expansion can be considered such that more derivatives of the density
are taken into account. Second, we can enforce a higher order
smoothness in the CR-algorithm.  Both methods are outlined below.

The proposed method can easily be extended by considering more terms
with higher order derivatives in the truncated series
\eqref{f_constant}. For example,
consider the expansion 
\begin{equation} \label{eq:higher_spatial}
  f=f^{eq}+\alpha\frac{\partial \rho}{\partial x}+\beta
  \frac{\partial^2 \rho}{\partial x^2}+\delta \frac{\partial^3
    \rho}{\partial x^3}+\epsilon \frac{\partial^4 \rho}{\partial x^4}
\end{equation}
that now requires the determination of more coefficients that are
found by considering --- in addition to $x_j$ and $x_k$ --- additional
grid points $x_l$ and $x_m$. This leads to a larger system of
unknowns but will give better results.

Higher order smoothness can be enforced on the moments $\phi$ and
$\xi$ as in the CR-algorithm by considering a higher order $m$ in
Eq.~\eqref{CR_formulas}. This requires more LBM steps and uses
a higher order finite difference formula to estimate the derivatives
in time.

For further conclusions and results higher order
derivatives and higher order smoothness are taken into account.




\subsection{Derivation macroscopic PDE}\label{der:PDE}
Next, we derive from Eq.~\eqref{constants}, the
macroscopic PDE by summing over the velocities. Using
$\sum_i f_i(x,t)=\rho(x,t)=\sum_i f_i^{eq}(x,t)$ results in a
macroscopic PDE for the density.
\begin{eqnarray*}
\left(\sum_i \gamma_i \right) \frac{\partial \rho}{\partial t} 
&+& \left(\sum_i \zeta_i \right) \frac{\partial^2 \rho}{\partial t^2}
 + \ldots + \left(\sum_i \eta_i \right) \frac{\partial^2 \rho}{\partial x \partial t} + \ldots \\
&= & - \left(\sum_i \alpha_i\right) \frac{\partial \rho}{\partial x} - \left(\sum_i \beta_i \right) \frac{\partial^2 \rho}{\partial x^2} - \left(\sum_i \delta_i \right)\frac{\partial^3 \rho}{\partial x^3}- \left(\sum_i \epsilon_i \right)\frac{\partial^4 \rho}{\partial x^4} - \ldots. \nonumber
\end{eqnarray*}

Series \eqref{constants} derived in this setting leads to the
classical Chapman-Enskog expansion \cite{leemput_phd, cercignani} that
we obtained in Section \ref{macro_LBM}. Indeed, Eq.~\eqref{constants} is written
as
\[
f_i(x,t)=g_i+\mathcal{L}_ig_i+\mathcal{L}_i^2g_i+\ldots,
\]
because of the application of the fixed point iteration.  When the
series is truncated after the second order spatial derivative and the
first order time derivative, we end up with
\[
f_i(x,t)=\frac{1}{3}\rho-\frac{i\Delta x}{3\omega}\frac{\partial \rho}{\partial x} + \bigg(\frac{i^2\Delta x^2}{3\omega^2}-\frac{i^2\Delta x^2}{6\omega}\bigg)\frac{\partial^2\rho}{\partial x^2}-\frac{\Delta t}{3\omega}\frac{\partial \rho}{\partial t}.
\]
Summing $f_i(x,t)$ over $i \in \{-1,0,1\}$ we obtain the same
macroscopic diffusion PDE \eqref{link_PDE_LBM}. Substituting this PDE
in the series to remove the time derivative leads to the classical
Chapman-Enskog expansion in Eq.~\eqref{analytical_chapman}.

This macroscopic PDE was used for the removal of the time derivative in
Eq.~\eqref{eq_coeff} and its replacement with
Eq.~\eqref{f_constant}.  Furthermore, it is important to
note that the macroscopic PDE is not necessarily of the
reaction-diffusion prototype.
Truncating \eqref{constants} after more terms and taking more
derivatives into account results in a better output but it will lead
to the term $\frac{\partial^2 \rho}{\partial t^2}$ which gives a less
comfortable macroscopic PDE.

This relation to the macroscopic PDE can now be integrated in the
numerical Chapman-Enskog method.  Once the fixed point described in
Section \ref{sec:distr_slow} is found, we have $\alpha$ and $\beta$
that lifts $\rho$ to the distribution functions close to the slow
manifold.  By performing two more LBM steps, $\frac{\partial
  \rho}{\partial t}$ can be calculated by using a forward finite
difference formula.  System \eqref{system_constants} can be applied to
find the vectors of constants of this larger system that include the time
derivative. There are now two possibilities: either the resulting
system is non-singular and it can be solved for the coefficients and
only an approximate PDE can be found as is considered above. Or it is
too singular to be solved accurately but then the PDE can be extracted
from the nullspace of the system.

Let us first discuss the situation where the matrix in
\eqref{system_constants} is non-singular. The system can then be
solved for $\alpha$, $\beta$ and $\gamma$. The approximate PDE can be
determined by summing over the obtained coefficients.
\[
\frac{\partial \rho}{\partial t}=-\frac{\sum_i \alpha_i }{\sum_i \gamma_i} \frac{\partial \rho}{\partial x} - \frac{\sum_i \beta_i}{\sum_i \gamma_i} \frac{\partial^2 \rho}{\partial x^2}.
\]
This PDE is only approximate. Otherwise, if it would hold exactly, the
system would be singular as expected.

For a singular system, we know that one or more of the eigenvalues
will be zero with a corresponding null eigenvector. Focusing on the
null eigenvector $v=\{v_1,v_2,\ldots, v_9\}$, we know that $Av=0$ with
$A$ the matrix in system \eqref{system_constants}. Using this, we
obtain
\[
\frac{\partial \rho(x_j)}{\partial x}v_1 + \frac{\partial^2 \rho(x_j)}{\partial x^2}v_4 + \frac{\partial \rho(x_j)}{\partial t}v_7 = 0,
\]
from which we conclude that the resulting PDE looks like
\[
\frac{\partial \rho(x_j)}{\partial t}=-\frac{v_1}{v_7} \frac{\partial \rho(x_j)}{\partial x} - \frac{v_4}{v_7}\frac{\partial^2 \rho(x_j)}{\partial x^2}.
\]
Remark that same PDE will be found when considering the equation in
grid points $x_k$ and $x_l$ instead of $x_j$.




\subsection{Algorithm for lifting operator and macroscopic PDE} \label{algorithm}
The results of the previous sections are now combined in an algorithm
that delivers a lifting operator and an approximate macroscopic
PDE. This can be used, for example, to construct a hybrid model. The
pseudocode is presented in Algorithm \ref{pseudocode} while the
complete algorithm is presented below. The algorithm starts by
searching for the lifting operator on the basis of the spatial
derivatives. Thereafter, it inserts time derivatives and calculates
the coefficients of the macroscopic PDE.

Start with an initial guess for $\{\alpha,\beta,\delta,\epsilon\}$ in
Eq.~\eqref{eq:higher_spatial}.  Apply Function \ref{fun:h}
$h(\alpha,\beta,\delta,\epsilon;\rho,m)$ for a given $\rho$ and a
certain $m$ for the order of smoothness.  This results in
coefficients $\{\alpha,\beta,\delta,\epsilon\}$ that represent
distribution functions closer to the slow manifold. The lifting
operator is constructed at this point.

When these distribution functions are found based on the spatial
derivatives only, we still need to determine the corresponding PDE by
considering the null eigenvector or by a summation of the coefficients as
discussed in Section \ref{der:PDE}.  By performing two extra
LBM steps --- to estimate the time derivative with a forward finite
difference formula --- the coefficient $\gamma$ belonging to the time
derivative of the expansion below can be numerically calculated.
\[
f=f^{eq}+\alpha \frac{\partial \rho}{\partial x}+\beta \frac{\partial^2 \rho}{\partial x^2}+\delta \frac{\partial^3 \rho}{\partial x^3}+\epsilon \frac{\partial^4 \rho}{\partial x^4}+\gamma \frac{\partial \rho}{\partial t}.
\]
Since the PDE can be obtained from the numerically constructed distribution functions, the PDE obtained through the Chapman-Enskog expansion does not need to be obtained analytically. 

\begin{remark}
Note that one can also consider $f^{eq}(x,t)=\kappa \rho(x,t)$ and determine the constants of vector $\kappa$ in a similar setting by using an extra grid point to obtain a larger system of unknowns.
\end{remark}
\begin{remark}
  In this paper we have chosen to use the same $\Delta x$ and
  $\Delta t$ in the LBM as in the PDE.  However, their stability
  properties may be different.  Our specific LBM simulation is stable in the
  2-norm when $0 \leq \omega \leq 2$ \cite{stability}. However, it is not necessary
  that the macroscopic PDE, when it is discretized with the same
  $\Delta x$ and $\Delta t$ and forward Euler, is also stable. Indeed,
  when $\omega \rightarrow 0$ for a fixed $\Delta x$ and $\Delta t$
  the resulting diffusion coefficient $D$ grows, see
  \eqref{link_PDE_LBM}, and this can lead to an instability.
\end{remark}

\begin{algorithm}
\caption{Pseudocode numerical Chapman-Enskog expansion} \label{pseudocode}
\begin{algorithmic}
\REQUIRE Test domain that defines $\rho(x,0)$, initial guess $a^0=\{\alpha,\beta,\delta,\epsilon\}$ = zeros(12,1) and a user-defined tolerance tol, parameter $m$ for higher order smoothness.
\REPEAT
\STATE $a^{k+1}=a^k-\bigl(J(a^k)\bigr)^{-1}h(a^k;\rho,m)$ with $h$ defined in Function \ref{fun:h}.
\UNTIL $\|a^{k+1}-a^k\|<$ tol.
\STATE Result for coefficients $\{\alpha,\beta,\delta,\epsilon\}$ belonging to the spatial derivatives.
\STATE Distribution functions $f=f^{eq}+\alpha \frac{ \partial \rho}{\partial x}+\beta \frac{\partial^2 \rho}{\partial x^2} + \delta \frac{\partial^3 \rho}{\partial x^3}+ \epsilon \frac{\partial^4 \rho}{\partial x^4}$ closer to the slow manifold.
\STATE Perform 2 more LBM steps to determine $\frac{\partial \rho}{\partial t}$ numerically by a forward finite difference formula.
\STATE Construct system \eqref{system_constants} --- including higher order spatial derivatives --- to achieve coefficients $\{\alpha,\beta,\delta,\epsilon,\gamma\}$.
\STATE Determine coefficients of PDE by using the nullspace of the system or by summation of the coefficients (Section \ref{der:PDE}). 
\RETURN Lifting operator $f=f^{eq}+\alpha \frac{ \partial \rho}{\partial x}+\beta \frac{\partial^2 \rho}{\partial x^2} + \delta \frac{\partial^3 \rho}{\partial x^3}+ \epsilon \frac{\partial^4 \rho}{\partial x^4}$ and macroscopic PDE.
\end{algorithmic}
\end{algorithm}

The pseudocode of the numerical Chapman-Enskog expansion as a lifting
operator is given in Algorithm \ref{pseudocode} together with the
determination of the transport coefficients of the PDE to construct a
hybrid model. The proposed algorithm has several advantages.  In
contrast to the Chapman-Enskog expansion no analytical work is
required.  Compared to the Constrained Runs algorithm it significantly
reduces the number of unknowns in the lifting since it only needs to
find the coefficients (vectors of constants) rather than the full
state of the distribution functions.  Furthermore, it can be done
off-line before the calculations. Indeed, once the coefficients are
found they can be reused every time step to realize the lifting. As an
extra surplus, the PDE can be determined to construct hybrid models.

%
%
\section{Numerical Results} \label{numerical_results}
The new lifting operator is now illustrated in several examples. First
we benchmark its accuracy against a reference solution that is
reconstructed. This is done in Section \ref{test_problem}. In Section
\ref{hybrid_1D} we recall the one-dimensional hybrid model of Figure
\ref{path}. Two-dimensional problems are discussed in Section
\ref{hybrid_2D}. The important comparison of the additional required
LBM steps to perform the lifting in a hybrid model is presented in
Section \ref{section_compare}.

\subsection{Numerical comparison of different lifting operators} 
\label{test_problem}

The proposed lifting operator can be tested against a reference
distribution function $f_c$.  This reference solution is calculated by
performing 1000 lattice Boltzmann steps starting from an initial state
that corresponds to the equilibrium distribution function of a given
density $\rho$.

The lifting operator can now be evaluated by restricting the reference
distribution function $f_c$ to its density $\rho = \sum_i f_i(x,t)$
and lift it back to a distribution function $f$ by using the proposed
lifting operator. The resulted $f$ will be compared with $f_{c}$ with
the help of the 2-norm $\|f-f_{c}\|$.

\begin{example} \label{ex:2} The considered model problem has the following parameters for a one-dimensional domain of length $L$.
\[
L = 10, \quad n = 200, \quad \Delta x = \frac{L}{n} = 0.05, \quad  \Delta t = 0.001, \quad \rho(x,0) = e^{-(x-\frac{L}{2})^2}, \quad \omega=0.9091.
\]
For these parameters the classical Chapman-Enskog expansion predicts a
diffusion coefficient $D=1$ (Eq.~\eqref{link_PDE_LBM}).
\end{example}


To reproduce the numerical results linked to Example \ref{ex:2} we include some extra information on how to determine the indices of \eqref{system_constants_spatial}. 
As mentioned in Section \ref{sec:distr_slow} we can choose an arbitrary initial density $\rho(x,0)$ and a test domain to determine the constants of the lifting operator \eqref{f_constant}. For example, consider $\rho(x,0)=x+1/2x^2$ for unknowns $\alpha, \beta$ and $\rho(x,0)= x+1/2x^2+1/6x^3$ for unknowns $\alpha, \beta, \delta$. $x$ is defined by the spatial nodes of the test domain defined below. Furthermore, the test domain reduces the computational expense compared to the actual spatial domain.

Consider the domain parameters of Example \ref{ex:2}. Since we know that the constants are only
  affected by these space and time steps, we should consider ---
  together with $\rho(x,0)$ --- a test domain with the
  same step sizes since the vectors of constants \eqref{vectors_constants} are affected by these choices. The test domain is of length $L_{test}=3$ such that $n_{test}=60$ since $\Delta x=0.05$. This number of grid points will make it possible to choose the indices $x_j$, $x_k$, $\ldots$ such that system
  \eqref{system_constants_spatial} is not close to singular. We can
  return now to the question which indices should be used
  in system \eqref{system_constants_spatial}. Focus on the fact that
  we do not want an effect of wrongly chosen boundary conditions in
  the smaller test domain. The grid points should be taken far enough
  from the edges and in points such that the system
  \eqref{system_constants_spatial} does not become singular. The indices can be, for example, $10$, $20$, $30$, $\ldots$ spread over the test domain of $60$ grid points.

  Note that one can also focus on local updates around the considered grid points $x_j$, $x_k$, $\ldots$.
  When the number of iterations needed in Newton's method are known, one knows how many LBM steps will be performed to find the new coefficients $\alpha$ and $\beta$. Then the size of the test domain can be shrunk to a smaller domain around $x_j$, $x_k$, $\ldots$. This is the same idea as used in \cite{vanderhoydonc} to perform local updates for the CR-algorithm.

To compare the proposed lifting operator with the existing ones discussed in Section \ref{review}, the results for $\|f-f_{c}\|$ of the different lifting operators are included in Tables \ref{compare_exact}, \ref{compare_exactCR} and \ref{compare_NCE}.

Table \ref{compare_exact} contains results obtained with the
analytical Chapman-Enskog expansion as a lifting operator. The first
column gives the order of the expansion and the second column shows
$\|f-f_{c}\|$. As expected a better accuracy is obtained with higher
order expansions. For example, by taking the third
derivative of the density into account an error of 1.24e-5 is achieved. 

\begin{table}[!htop]
\caption{The error $\|f-f_c\|$ is presented to test the exact Chapman-Enskog expansion as a lifting operator. The reference solution is obtained after 1000 LBM steps in the model problem from Example \ref{ex:2}. \label{compare_exact}}
\begin{center} \footnotesize
\begin{tabular}{|l | l|}\hline
Construction lifting operator: & $\|f-f_c\|$  \\
exact Chapman-Enskog expansion &  \\ \hline
$f=f^{eq}$  &  0.0388 \\
$f=f^{eq}+\alpha_{exact} \frac{\partial \rho}{\partial x}$&   5.2341e-004 \\
$f=f^{eq}+\alpha_{exact} \frac{\partial \rho}{\partial x}+\beta_{exact} \frac{\partial \rho^2}{\partial x^2}$  &  2.7570e-005 \\
$f=f^{eq}+\alpha_{exact} \frac{\partial \rho}{\partial x}+\beta_{exact} \frac{\partial \rho^2}{\partial x^2}+\delta_{exact} \frac{\partial^3 \rho}{\partial x^3}$ &  1.2439e-005 \\ \hline
\end{tabular}
\end{center}
\end{table}

In Table \ref{compare_exactCR} we
use the Constrained Runs algorithm of various orders of accuracy to
numerically lift the density to distribution functions.  Different
types of backward extrapolation are listed in the first column of the
table while the corresponding 2-norm $\|f-f_c\|$ is described in the
second column.  There we see that very accurate results can be found
for the higher order versions.  Note that these methods find for each
grid point the moments $\phi$ and $\xi$ of the distribution
functions. Together with $\rho$, the corresponding distribution
functions are found by Eq.~\eqref{omzet}. Since this gives a local
solution it can give accurate results. The last column contains $\|f-f_c\|$
when some extra advection effect is included, which shows similar results as the pure
diffusion problem.

\begin{table}[!htop]
  \caption{The error $\|f-f_c\|$ with the Constrained Runs algorithm (combined with Newton's method) for various orders of accuracy as a lifting operator. The reference solution is obtained for the model problem in Example \ref{ex:2} by performing 1000 LBM time steps before restricting and lifting. The last column contains results when an extra advection effect of $a=0.66$ is included --- which changes the used equilibrium distribution functions as noted in Eq.~\eqref{eq:advection}.
\label{compare_exactCR}}
\begin{center} \footnotesize
\begin{tabular}{|l | l| l|}\hline
Construction lifting operator: & $\|f-f_c\|$ & $\|f-f_c\|$\\
extrapolation CR-algorithm  & pure diffusion $D=1$ & plus advection $a=0.66$\\ \hline
Constant  &  0.0010 & 0.0014\\
Linear &  1.3578e-006  & 1.7927e-006\\
Quadratic  &  2.9359e-009 & 3.9069e-009\\
Cubic &  9.0125e-012 & 1.1898e-011\\ \hline
\end{tabular}
\end{center}
\end{table}

Table \ref{compare_NCE} shows the results with the proposed numerical Chapman-Enskog expansion as a lifting operator.  We clearly see that taking more terms
  in the expansion, i.e.~more derivatives of $\rho(x,t)$ in the
  lifting, leads to a better lifting operator.   In the same table we show the results with
  higher order smoothness conditions by using Eq.~\eqref{CR_formulas} with higher order $m$. As in the CR-algorithm, higher order smoothness does result in a significant improvement.
  The accuracy increases
  to 9.45e-11 when up to the sixth spatial derivative is
  taken into account. Including advection in this table will also show similar results but these are not added.

\begin{table}[!htop]
  \caption{
    The error $\|f-f_c\|$ with the numerical Chapman-Enskog expansion as a lifting operator.  
    The reference distribution function $f_c$ is obtained by performing 1000 LBM time steps with parameters listed in
    Example \ref{ex:2}.  Each of the lifting operators is calculated
    as a fixed point for the coefficients. 
    The first table shows the results with one LBM step before updating the moments, implying an update of the coefficients. 
    There are results listed for increasing number of terms in the expansion, implying an increasing number of considered coefficients. 
    The second table shows the same results where two LBM steps are used to estimate the smoothness.
    The third and fourth table show the results with a quadratic, respectively cubic computation of the finite difference approximation in Eq.~\eqref{CR_formulas}. The constant coefficients found via the numerical procedure are compared to those found exactly with the classical Chapman-Enskog expansion in columns 3, 4 and 5.
\label{compare_NCE}}
 \begin{center} \footnotesize
\begin{tabular}{|l | l| l| l| l|}\hline
Construction lifting operator: & $\|f-f_c\|$ & $\| \alpha - \alpha_{exact}\|$ &  $\| \beta - \beta_{exact}\|$ & $\| \delta -\delta_{exact} \|$\\
Numerical Chapman-Enskog & & & & \\
constant computation of fixed point  & & & & \\
\hline
$f=f^{eq}+\alpha \frac{\partial \rho}{\partial x}$& 5.2341e-004 &  1.9981e-016 & /& /  \\
$f=f^{eq}+\alpha \frac{\partial \rho}{\partial x}+\beta \frac{\partial \rho^2}{\partial x^2}$  &    0.0010 &  3.7153e-016 &     8.9815e-004 & / \\
$f=f^{eq}+\alpha \frac{\partial \rho}{\partial x}+\beta \frac{\partial \rho^2}{\partial x^2}+\delta \frac{\partial^3 \rho}{\partial x^3}$ &   0.0010  &     2.0200e-015 &   8.9815e-004 &   2.0310e-005 \\
$f=f^{eq}+\alpha \frac{\partial \rho}{\partial x}+\beta \frac{\partial \rho^2}{\partial x^2}+\delta \frac{\partial^3 \rho}{\partial x^3}+\epsilon \frac{\partial^4 \rho}{\partial x^4}$ &   0.0010 &     5.0363e-015
 &    8.9815e-004 &   2.0310e-005 \\ 
 $f=f^{eq}+\alpha \frac{\partial \rho}{\partial x}+\beta \frac{\partial \rho^2}{\partial x^2}+\delta \frac{\partial^3 \rho}{\partial x^3}+\epsilon \frac{\partial^4 \rho}{\partial x^4} + \theta \frac{\partial^5 \rho}{\partial x^5} $ &  0.0010 &       1.2979e-013 &     8.9815e-004 &   2.0310e-005 \\ 
$f=f^{eq}+\alpha \frac{\partial \rho}{\partial x}+\beta \frac{\partial \rho^2}{\partial x^2}+\delta \frac{\partial^3 \rho}{\partial x^3}+\epsilon \frac{\partial^4 \rho}{\partial x^4} +\theta \frac{\partial^5 \rho}{\partial x^5}+ \iota \frac{\partial^6 \rho}{\partial x^6}$ &  0.0010 &     9.0311e-013 &    8.9815e-004 &    2.0310e-005 \\ \hline
\end{tabular}
\begin{tabular}{|l | l| l| l| l|}\hline
linear computation of fixed point & $\|f-f_c\|$ & $\| \alpha - \alpha_{exact}\|$ &  $\| \beta - \beta_{exact}\|$ & $\| \delta -\delta_{exact} \|$\\
\hline
$f=f^{eq}+\alpha \frac{\partial \rho}{\partial x}$& 5.2341e-004 & 9.0005e-017 & / & / \\
$f=f^{eq}+\alpha \frac{\partial \rho}{\partial x}+\beta \frac{\partial \rho^2}{\partial x^2}$  &  2.7570e-005 & 2.3990e-015 & 5.3202e-015 & /
 \\ 
$f=f^{eq}+\alpha \frac{\partial \rho}{\partial x}+\beta \frac{\partial \rho^2}{\partial x^2}+\delta \frac{\partial^3 \rho}{\partial x^3}$ &  6.8677e-007  &  9.1594e-015  &   2.1865e-014 &  1.0803e-005 \\
$f=f^{eq}+\alpha \frac{\partial \rho}{\partial x}+\beta \frac{\partial \rho^2}{\partial x^2}+\delta \frac{\partial^3 \rho}{\partial x^3}+\epsilon \frac{\partial^4 \rho}{\partial x^4}$ &  1.3591e-006 &    7.2493e-015 &   1.6524e-014 &   1.0803e-005 \\ 
$f=f^{eq}+\alpha \frac{\partial \rho}{\partial x}+\beta \frac{\partial \rho^2}{\partial x^2}+\delta \frac{\partial^3 \rho}{\partial x^3}+\epsilon \frac{\partial^4 \rho}{\partial x^4} + \theta \frac{\partial^5 \rho}{\partial x^5} $ &  1.3591e-006 &      1.4579e-013 &     3.6404e-013 &   1.0803e-005 \\ 
$f=f^{eq}+\alpha \frac{\partial \rho}{\partial x}+\beta \frac{\partial \rho^2}{\partial x^2}+\delta \frac{\partial^3 \rho}{\partial x^3}+\epsilon \frac{\partial^4 \rho}{\partial x^4} +\theta \frac{\partial^5 \rho}{\partial x^5}+ \iota \frac{\partial^6 \rho}{\partial x^6}$ &  1.3578e-006 &      2.6277e-012 &    6.4797e-012 &   1.0803e-005 \\ \hline
\end{tabular}
\begin{tabular}{|l | l| l| l| l|}\hline
quadratic computation of fixed point & $\|f-f_c\|$ & $\| \alpha - \alpha_{exact}\|$ &  $\| \beta - \beta_{exact}\|$ & $\| \delta -\delta_{exact} \|$\\
\hline
$f=f^{eq}+\alpha \frac{\partial \rho}{\partial x}$&  5.2341e-004 & 1.2647e-016 & /& /  \\
$f=f^{eq}+\alpha \frac{\partial \rho}{\partial x}+\beta \frac{\partial \rho^2}{\partial x^2}$  &   2.7570e-005 &  7.2802e-016 &
  1.1928e-015 & / \\
$f=f^{eq}+\alpha \frac{\partial \rho}{\partial x}+\beta \frac{\partial \rho^2}{\partial x^2}+\delta \frac{\partial^3 \rho}{\partial x^3}$ &    6.8677e-007  &   1.1455e-014 &   2.7241e-014 &   1.0803e-005 \\
$f=f^{eq}+\alpha \frac{\partial \rho}{\partial x}+\beta \frac{\partial \rho^2}{\partial x^2}+\delta \frac{\partial^3 \rho}{\partial x^3}+\epsilon \frac{\partial^4 \rho}{\partial x^4}$ &  4.1047e-008 &    2.9450e-014 &   6.4892e-014 &  1.0803e-005 \\ 
$f=f^{eq}+\alpha \frac{\partial \rho}{\partial x}+\beta \frac{\partial \rho^2}{\partial x^2}+\delta \frac{\partial^3 \rho}{\partial x^3}+\epsilon \frac{\partial^4 \rho}{\partial x^4} + \theta \frac{\partial^5 \rho}{\partial x^5} $ & 1.4994e-009 &     3.3586e-013 &    8.3590e-013 &   1.0803e-005 \\ 
$f=f^{eq}+\alpha \frac{\partial \rho}{\partial x}+\beta \frac{\partial \rho^2}{\partial x^2}+\delta \frac{\partial^3 \rho}{\partial x^3}+\epsilon \frac{\partial^4 \rho}{\partial x^4} +\theta \frac{\partial^5 \rho}{\partial x^5}+ \iota \frac{\partial^6 \rho}{\partial x^6}$ & 2.9449e-009 &      1.3229e-011 &   3.3252e-011 &   1.0803e-005 \\ \hline
\end{tabular}
\begin{tabular}{|l | l| l| l| l|}\hline
cubic computation of fixed point & $\|f-f_c\|$ & $\| \alpha - \alpha_{exact}\|$ &  $\| \beta - \beta_{exact}\|$ & $\| \delta -\delta_{exact} \|$\\
\hline
$f=f^{eq}+\alpha \frac{\partial \rho}{\partial x}$&   5.2341e-004 & 7.9768e-016 & /& /  \\
$f=f^{eq}+\alpha \frac{\partial \rho}{\partial x}+\beta \frac{\partial \rho^2}{\partial x^2}$  &   2.7570e-005 &   4.1708e-015 &
  1.0779e-014 & / \\
$f=f^{eq}+\alpha \frac{\partial \rho}{\partial x}+\beta \frac{\partial \rho^2}{\partial x^2}+\delta \frac{\partial^3 \rho}{\partial x^3}$ &   6.8677e-007  &   1.2825e-014 &    3.5932e-014 &   1.0803e-005 \\
$f=f^{eq}+\alpha \frac{\partial \rho}{\partial x}+\beta \frac{\partial \rho^2}{\partial x^2}+\delta \frac{\partial^3 \rho}{\partial x^3}+\epsilon \frac{\partial^4 \rho}{\partial x^4}$ & 4.1047e-008 &    3.9984e-014 &  7.9750e-014 &   1.0803e-005\\ 
$f=f^{eq}+\alpha \frac{\partial \rho}{\partial x}+\beta \frac{\partial \rho^2}{\partial x^2}+\delta \frac{\partial^3 \rho}{\partial x^3}+\epsilon \frac{\partial^4 \rho}{\partial x^4} + \theta \frac{\partial^5 \rho}{\partial x^5} $ & 1.4995e-009 &    6.9980e-013 &   1.6787e-012 &   1.0803e-005\\ 
$f=f^{eq}+\alpha \frac{\partial \rho}{\partial x}+\beta \frac{\partial \rho^2}{\partial x^2}+\delta \frac{\partial^3 \rho}{\partial x^3}+\epsilon \frac{\partial^4 \rho}{\partial x^4} +\theta \frac{\partial^5 \rho}{\partial x^5}+ \iota \frac{\partial^6 \rho}{\partial x^6}$ & 9.4492e-011 &     2.2682e-011 &   5.5927e-011 &   1.0803e-005\\ \hline
\end{tabular}
\end{center}
\end{table}

A comparison of Tables \ref{compare_exact},
  \ref{compare_exactCR} and \ref{compare_NCE} shows
that the proposed numerical lifting operator leads to better results
than the analytically found Chapman-Enskog expansion. As can be seen is the Constrained Runs algorithm a good lifting method, but, as
will be discussed in Section \ref{section_compare}, the computational expense of this method brings down
the beauty of it.  Table \ref{compare_LBM} of Section \ref{section_compare} contains a comparison
of the number of additional LBM steps required for each
of the lifting operators. In the CR-algorithm this additional cost
can be attributed to the construction of the Jacobian matrix. These
additional LBM steps make the method computationally very expensive.
In two dimensions this method becomes prohibitive.  This makes the numerical
Chapman-Enskog lifting operator a good alternative to the CR-algorithm that gives
a similar accuracy at a limited computational cost.


\subsection{One-dimensional test problem} \label{hybrid_1D} 
\begin{figure}[!htop]
\begin{center}
{\includegraphics[width=0.48\textwidth]{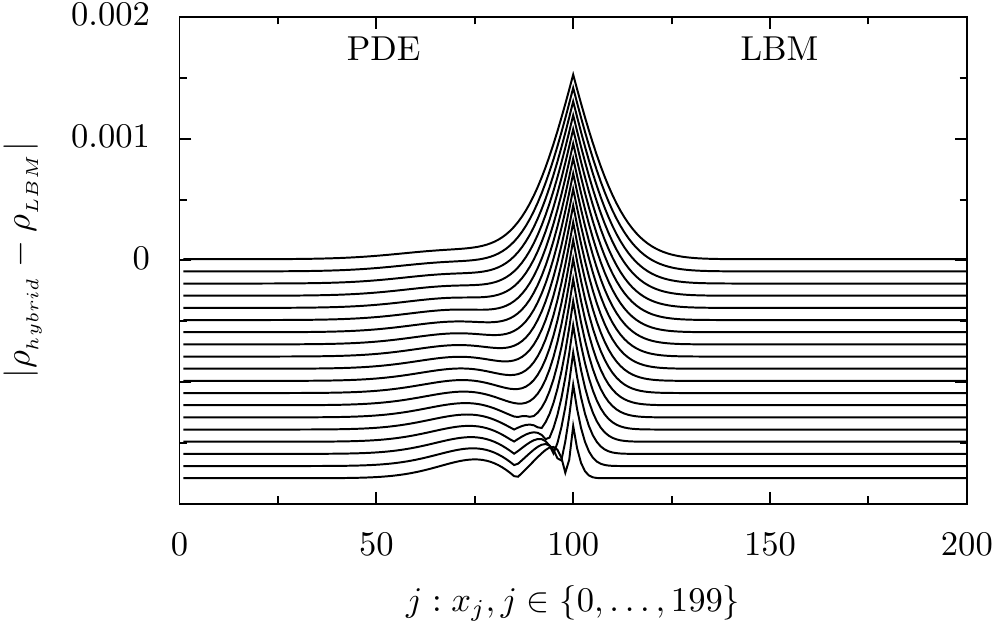}}{\includegraphics[width=0.5\textwidth]{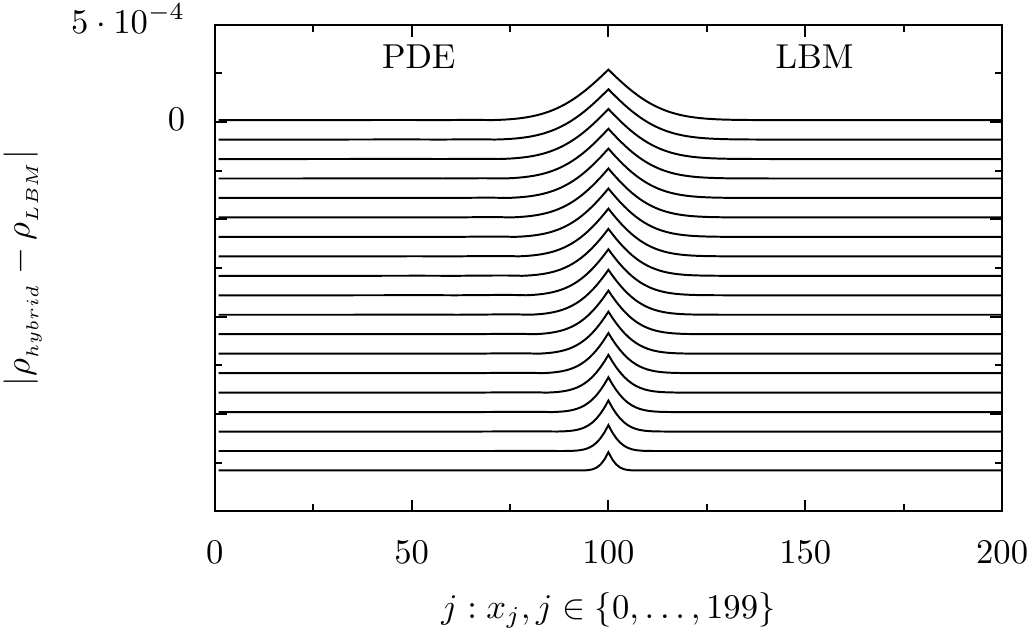}}\\
{\includegraphics[width=0.5\textwidth]{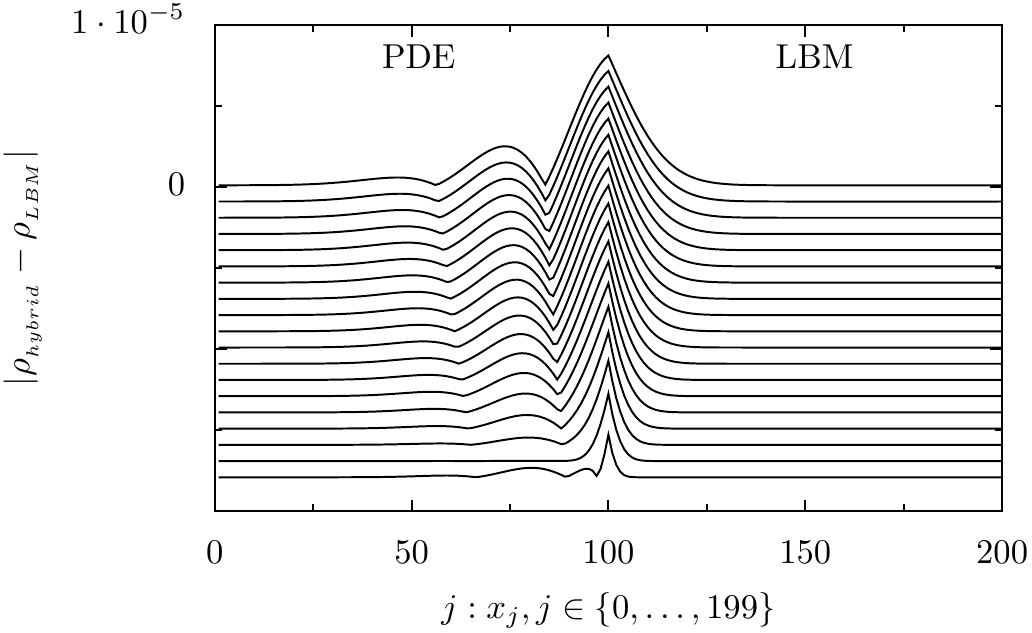}}{\includegraphics[width=0.5\textwidth]{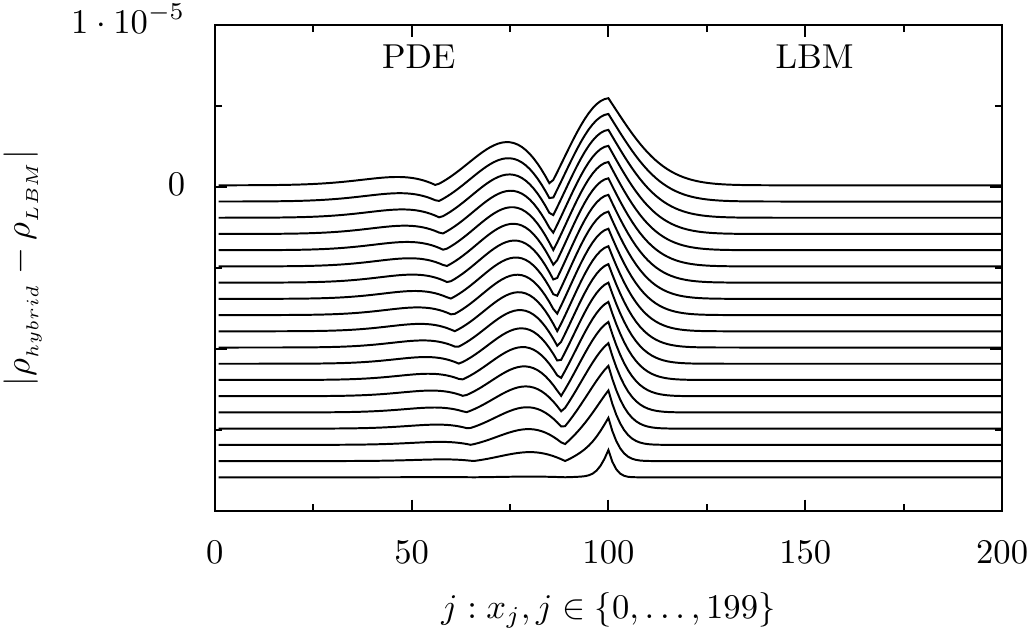}}
\end{center}
\caption{$|\rho_{_{\text{hybrid}}}-\rho_{_{\text{LBM}}}|$ after 200 time steps.  The
  difference is also shown at earlier time slots, but shifted down for
  clarity. The lines represent time steps between one and 200. The top
  line corresponds to time step 191 while the bottom line represents
  time step 11. The lines in between correspond to jumps with 10 time
  steps from 11 to 21, $\ldots$,181 to 191.  The domain that is
  considered, is shown in Figure \ref{path}. The lifting operator is $f_i(x,t)
  = 1/3 \, \rho(x,t)$ (top left), first order (top right), second order (bottom left) and third order
  Chapman-Enskog (bottom right) respectively. The lifting operator is used
  both to find the ghost points of the LBM domain and for the creation
  of the initial state for the LBM region. The model problem parameters are listed in Example \ref{ex:2}.
 \label{figures_exact}}
\end{figure}

\begin{figure}[!htop]
\begin{center}
\includegraphics[width=0.5\textwidth]{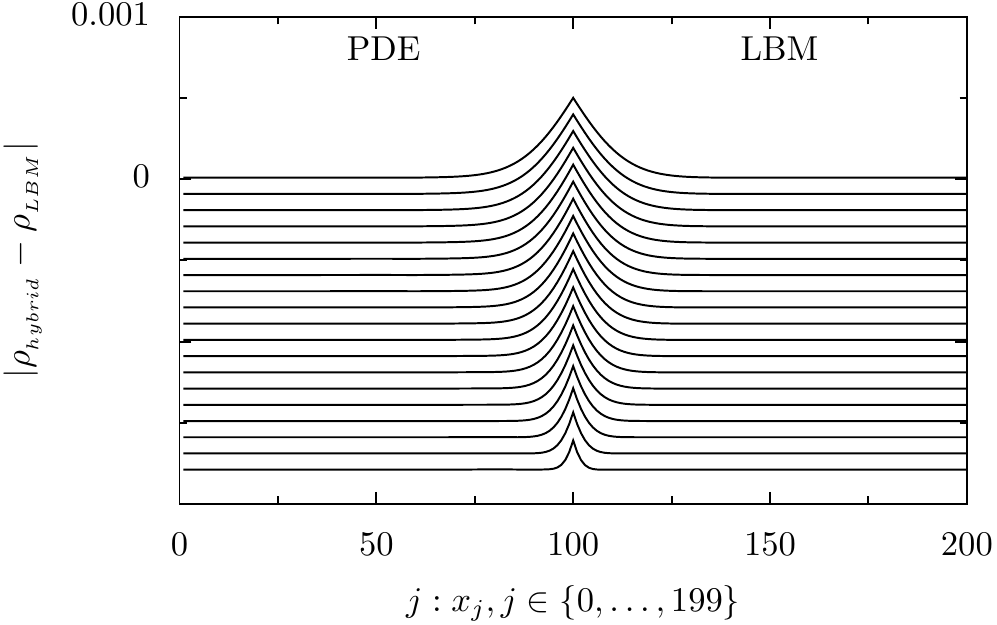}\includegraphics[width=0.5\textwidth]{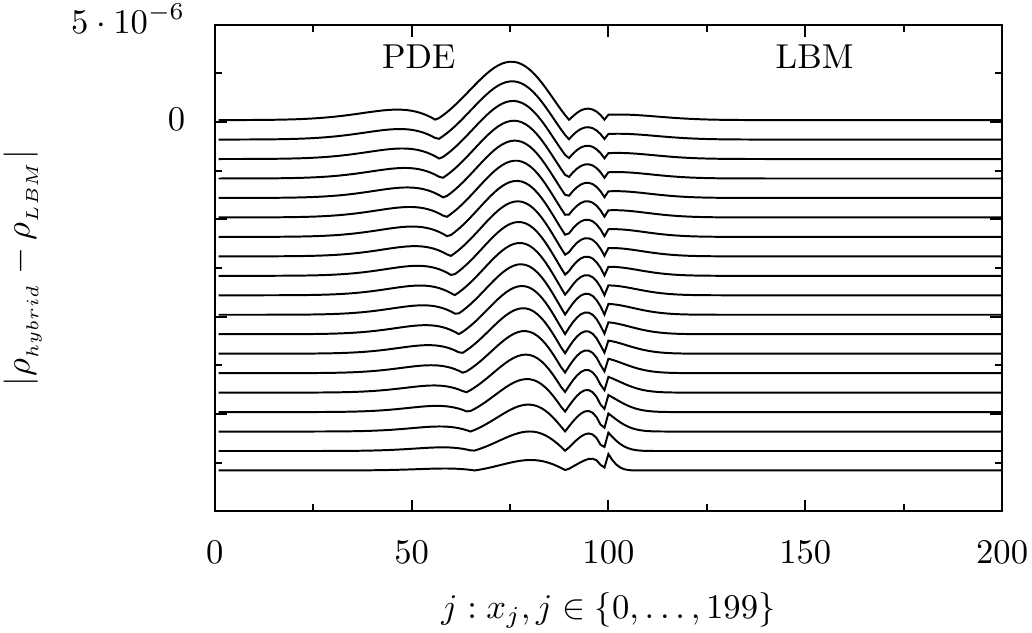}\\
\includegraphics[width=0.5\textwidth]{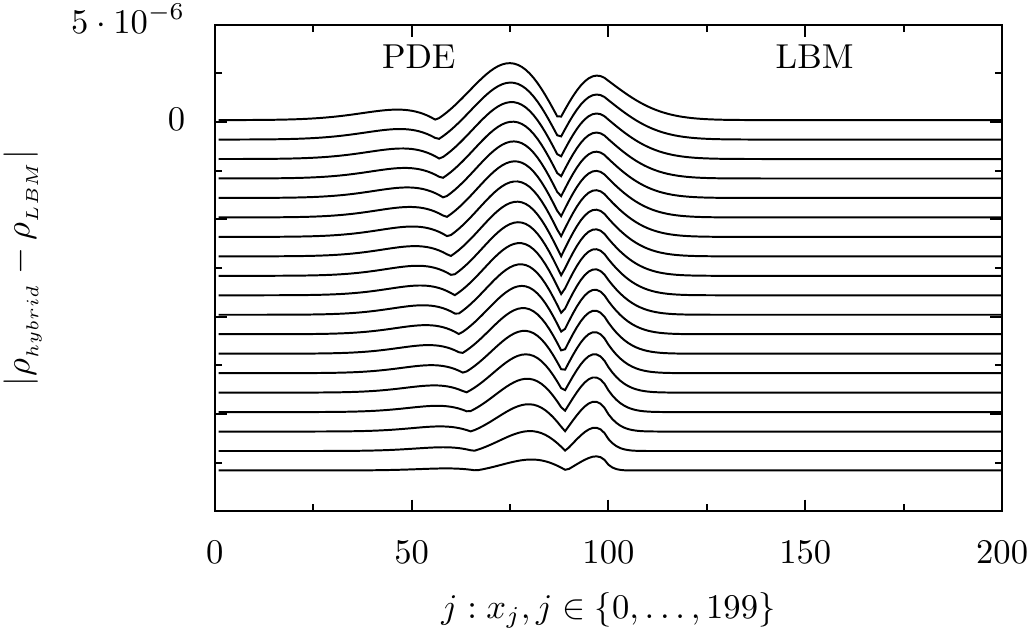}\includegraphics[width=0.5\textwidth]{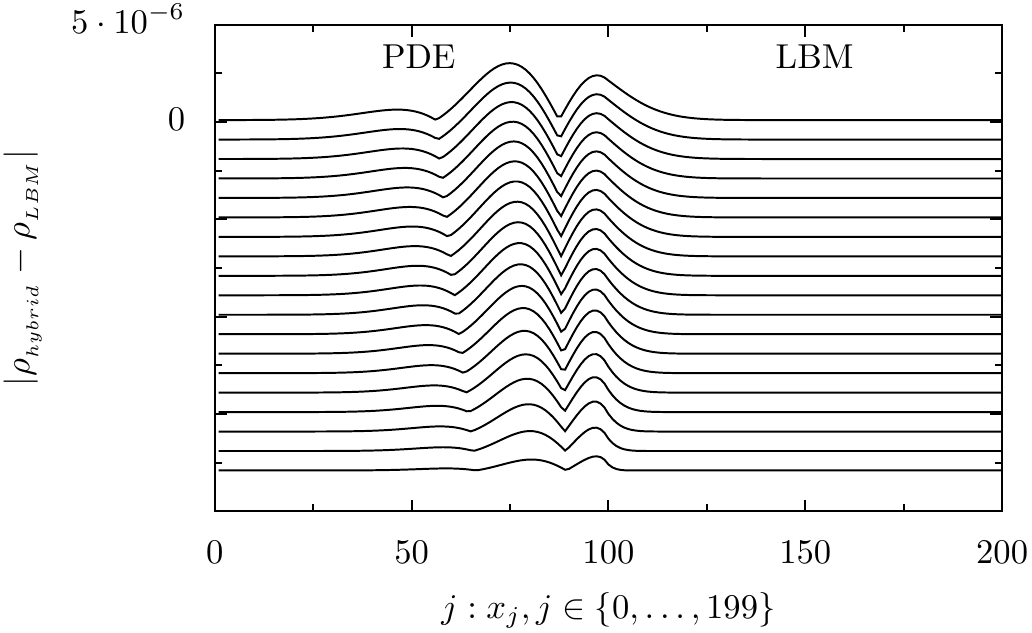}

\caption{$|\rho_{_{\text{hybrid}}}-\rho_{_{\text{LBM}}}|$ after 200
  time steps where the lifting operator based on the CR-algorithm is
  used in combination with the method of Newton.  We show results for
  constant (top left), linear (top right), quadratic (bottom left) and cubic (bottom right) backward
  extrapolation, respectively. The difference is also shown at
  earlier time slots, but shifted down for clarity. The domain that is
  considered, is shown in Figure \ref{path}. The model problem parameters are listed in Example \ref{ex:2}. \label{figures_CR}}
\end{center}
\end{figure}

To compare
the results of the numerical Chapman-Enskog expansion with the earlier proposed
lifting operators discussed in Section \ref{review}, Figures \ref{figures_exact} and \ref{figures_CR} show the
results of the absolute difference
$|\rho_{_{\text{hybrid}}}-\rho_{_{\text{LBM}}}|$ for the exact Chapman-Enskog
expansion and those obtained with the Constrained Runs
algorithm. $\rho_{_{\text{hybrid}}}$ is the density of the hybrid model and $\rho_{_{\text{LBM}}}$ the density of a full
LBM. $\rho_{_{\text{LBM}}}$ is the reference solution to compare the hybrid solution with. It considers a LBM on the whole spatial domain $[a,b]$ with the parameters outlined in Example \ref{ex:2} and the domain represented in Figure \ref{path}. The lifting operators are used both to initialize the LBM and to find the ghost points of the LBM domain.
Figure \ref{figures_exact} shows the absolute differences
by using as lifting operator the exact Chapman-Enskog expansion respectively
up to zeroth (top left), first (top right), second (bottom left) and third order (bottom right). Figure
\ref{figures_CR} shows the absolute differences with the lifting operator
based on the Constrained Runs algorithm in combination with Newton's
method for respectively a constant (top left), linear (top right), quadratic (bottom left) and cubic
(bottom right) extrapolation in time. These results were obtained in
\cite{vanderhoydonc} by considering local updates at the ghost points
of the LBM domain.

\begin{figure}[!htop]
\begin{center}
\includegraphics[width=0.5\textwidth]{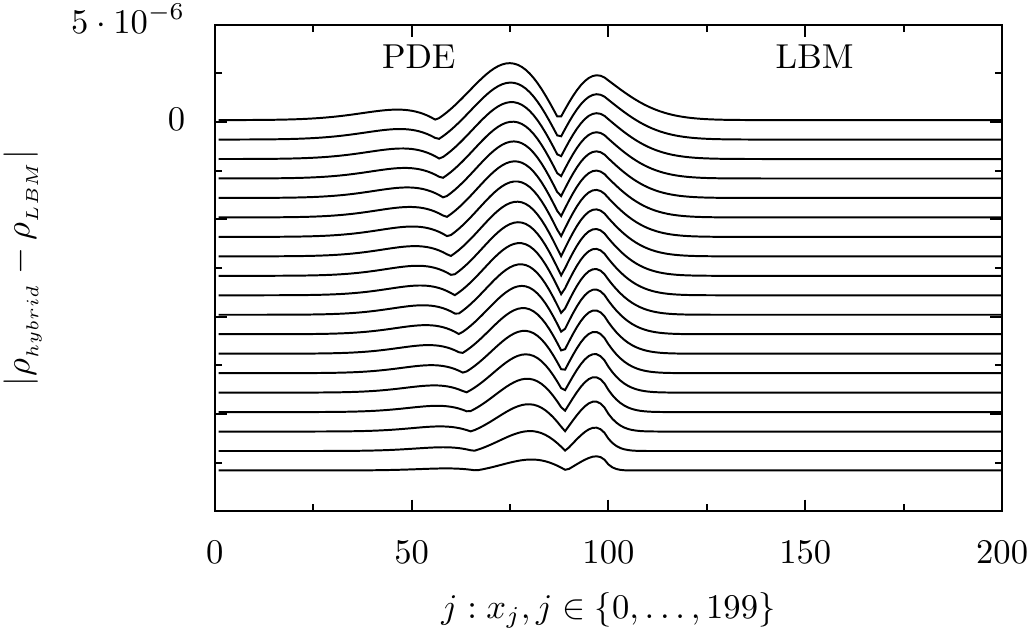}\includegraphics[width=0.5\textwidth]{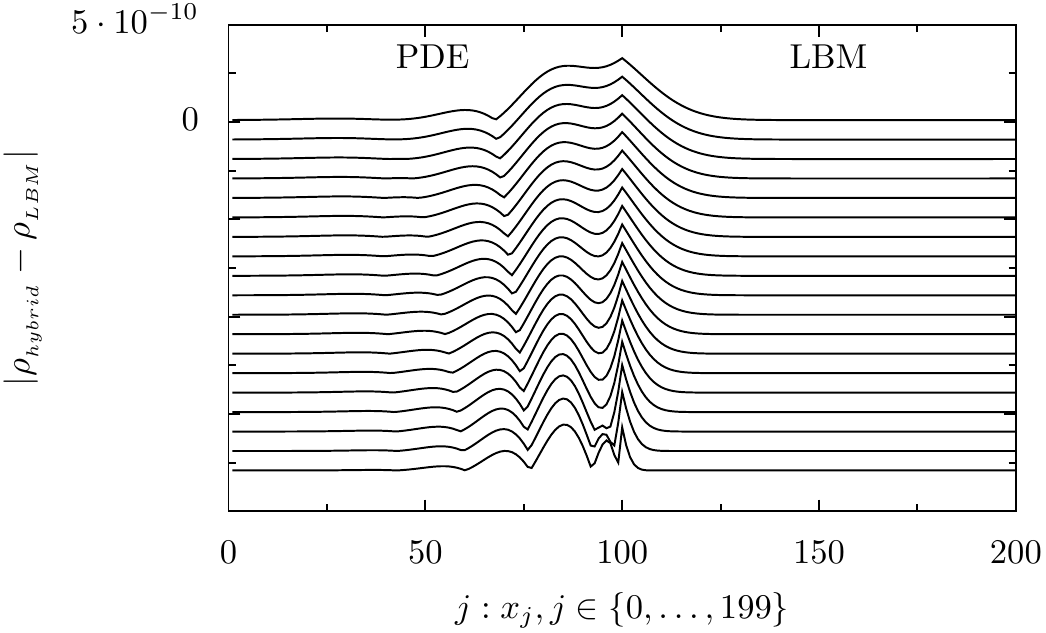}
\caption{$|\rho_{_{\text{hybrid}}}-\rho_{_{\text{LBM}}}|$ after 200 time steps in the model problem of Example \ref{ex:2}. The domain that is considered, is shown in Figure \ref{path}. To deal with the initial error and the error in the ghost points of the LBM domain the numerical Chapman-Enskog expansion (order spatial expansion 6) is used. The considered PDE in the hybrid domain is the analytically known PDE \eqref{link_PDE_LBM} in the left figure and the one that is obtained from the numerical Chapman-Enskog expansion in the right figure. \label{temepsilon}}
\end{center}
\end{figure}

When the numerical Chapman-Enskog expansion (up to the sixth spatial derivative) is used in our one-dimensional hybrid model problem, Figure \ref{temepsilon} is obtained. Here, we have two possibilities. First, act as if we know the PDE \eqref{link_PDE_LBM} obtained from the exact Chapman-Enskog expansion. $|\rho_{_{\text{hybrid}}}-\rho_{_{\text{LBM}}}|$ is given in the left Figure \ref{temepsilon} for which the hybrid domain is shown in Figure \ref{path} and the PDE is the one given in Eq.~\eqref{link_PDE_LBM}. Second, use the PDE that is obtained from the proposed lifting operator through summing the proposed lifting operator or considering the nullspace as explained in Section \ref{der:PDE}. With this PDE, the result for $|\rho_{_{\text{hybrid}}}-\rho_{_{\text{LBM}}}|$ is shown in the right Figure \ref{temepsilon}.

As can be seen in Figure \ref{temepsilon}, a change in the PDE --- by considering the PDE obtained through the numerical Chapman-Enskog expansion --- results in an even smaller modeling error compared to the one obtained via the classical Chapman-Enskog expansion.

\begin{figure}[!htop]
\begin{center}
\includegraphics[width=0.48\textwidth]{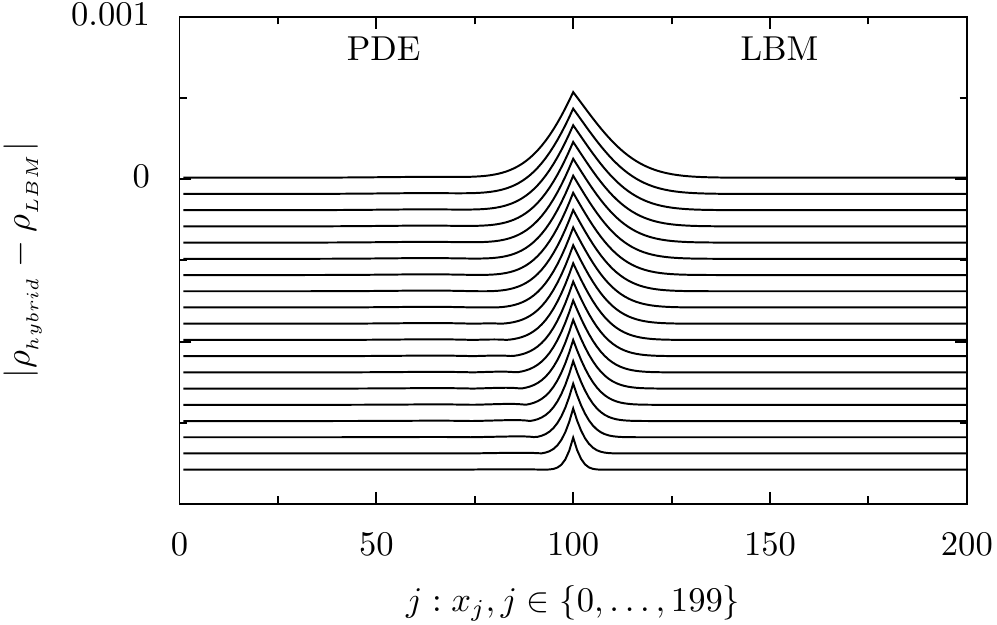}\includegraphics[width=0.5\textwidth]{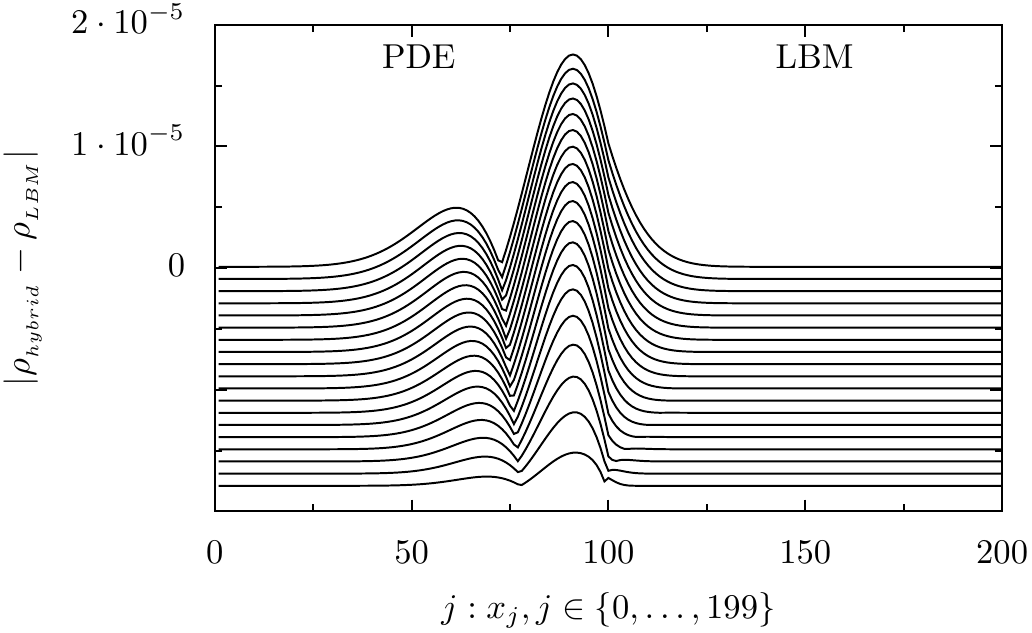}\\
\includegraphics[width=0.5\textwidth]{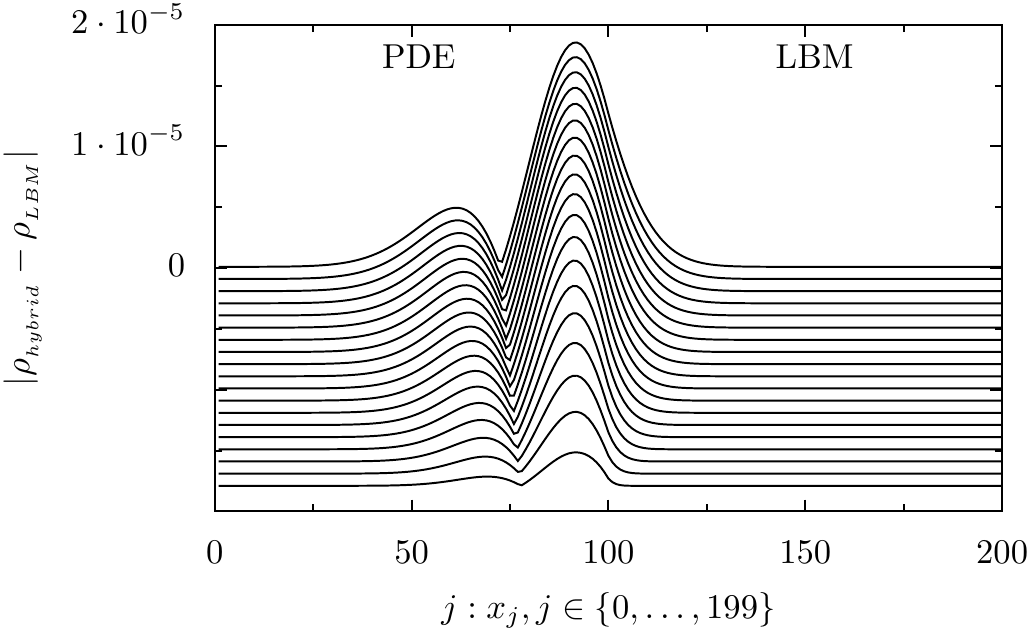}\includegraphics[width=0.5\textwidth]{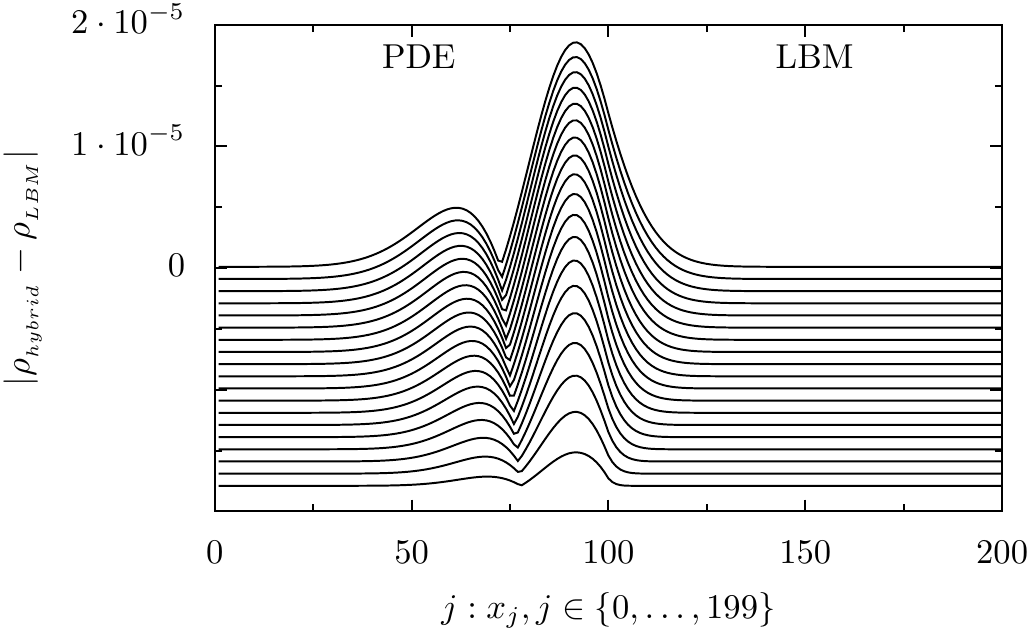}
\caption{$|\rho_{_{\text{hybrid}}}-\rho_{_{\text{LBM}}}|$ after 200
  time steps where the lifting operator based on the CR-algorithm is
  used in combination with the method of Newton.  We show results for
  constant (top left), linear (top right), quadratic (bottom left) and cubic (bottom right) backward
  extrapolation, respectively. The difference is also shown at
  earlier time slots, but shifted down for clarity. The domain that is
  considered, is shown in Figure \ref{path}. Model parameters are listed in Example \ref{ex:2} with an extra advection coefficient $a=0.66$. \label{CR_advectie}}
\end{center}
\end{figure}

\begin{figure}[!htop]
\begin{center}
\includegraphics[width=0.5\textwidth]{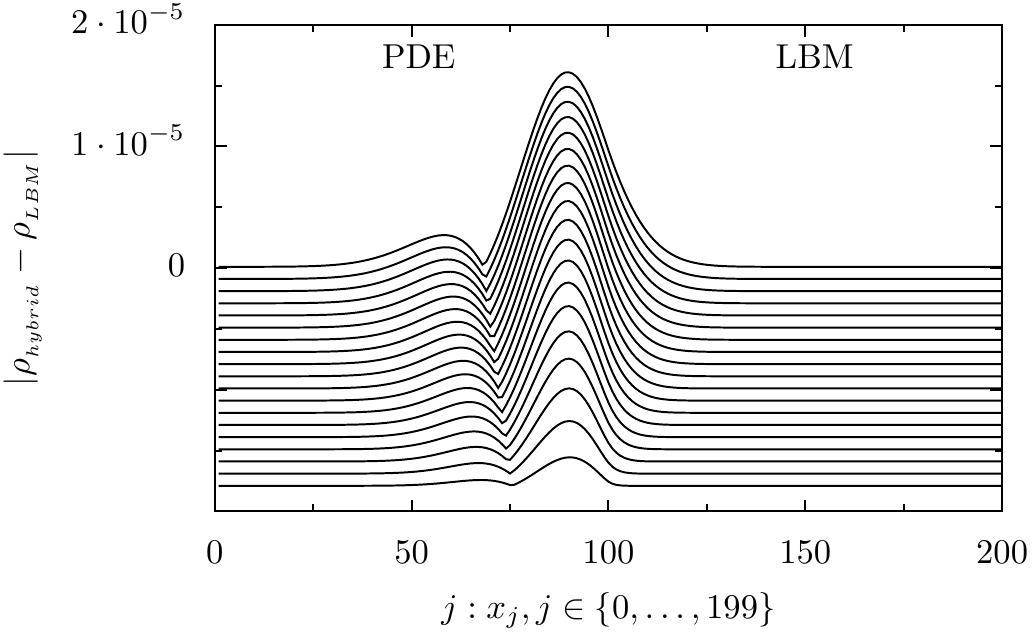}\includegraphics[width=0.5\textwidth]{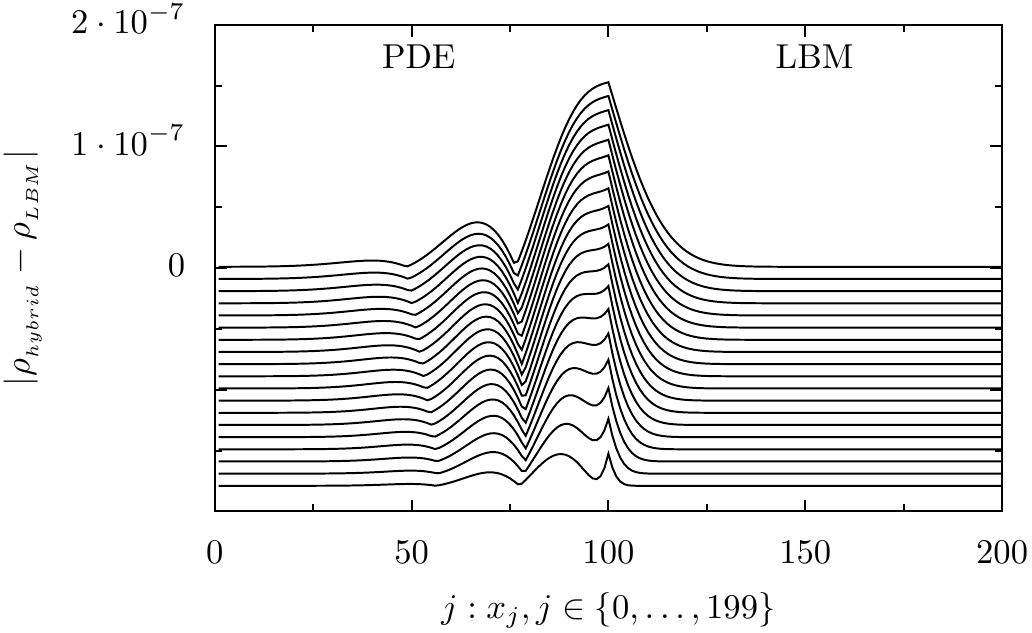}
\caption{$|\rho_{_{\text{hybrid}}}-\rho_{_{\text{LBM}}}|$ after 200 time steps in the model problem of Example \ref{ex:2} with advection effect $a=0.66$.
The domain that is considered, is shown in Figure \ref{path}. To deal with the initial error and the error in the ghost points of the LBM domain the numerical Chapman-Enskog expansion (order spatial expansion 4) is used. The considered PDE in the hybrid domain is the analytically known PDE \eqref{link_advection_LBM} in the left figure and the one that is obtained from the numerical Chapman-Enskog expansion in the right figure. \label{advectie}}
\end{center}
\end{figure}

Changing the parameters of the model such that advection is included,
is considered below.  The figures show similar results with
advection-term $a=0.66$. The model problem remains the one from Example \ref{ex:2}. The only difference is the change in the equilibrium distribution as shown in \eqref{eq:advection}.
Figure \ref{CR_advectie} contains the comparison results
obtained through the CR-algorithm. Figure \ref{advectie} (left) shows
$|\rho_{_{\text{hybrid}}}-\rho_{_{\text{LBM}}}|$ with the numerical Chapman-Enskog expansion as a lifting operator and the PDE obtained through the
Chapman-Enskog expansion while Figure \ref{advectie} (right) uses
the PDE obtained from the proposed lifting operator.

\subsection{Two-dimensional test problem} \label{hybrid_2D}
This section generalizes the previous one. Two spatial dimensions are considered.
Two-dimensional problems can take different discrete sets of velocities into account.
In Section \ref{D2Q5} results for D2Q5 are presented while Section \ref{D2Q9} contains results for D2Q9.

\subsubsection{D2Q5} \label{D2Q5}

\begin{figure}[!htop]
\begin{center}
\includegraphics[width=0.6\textwidth]{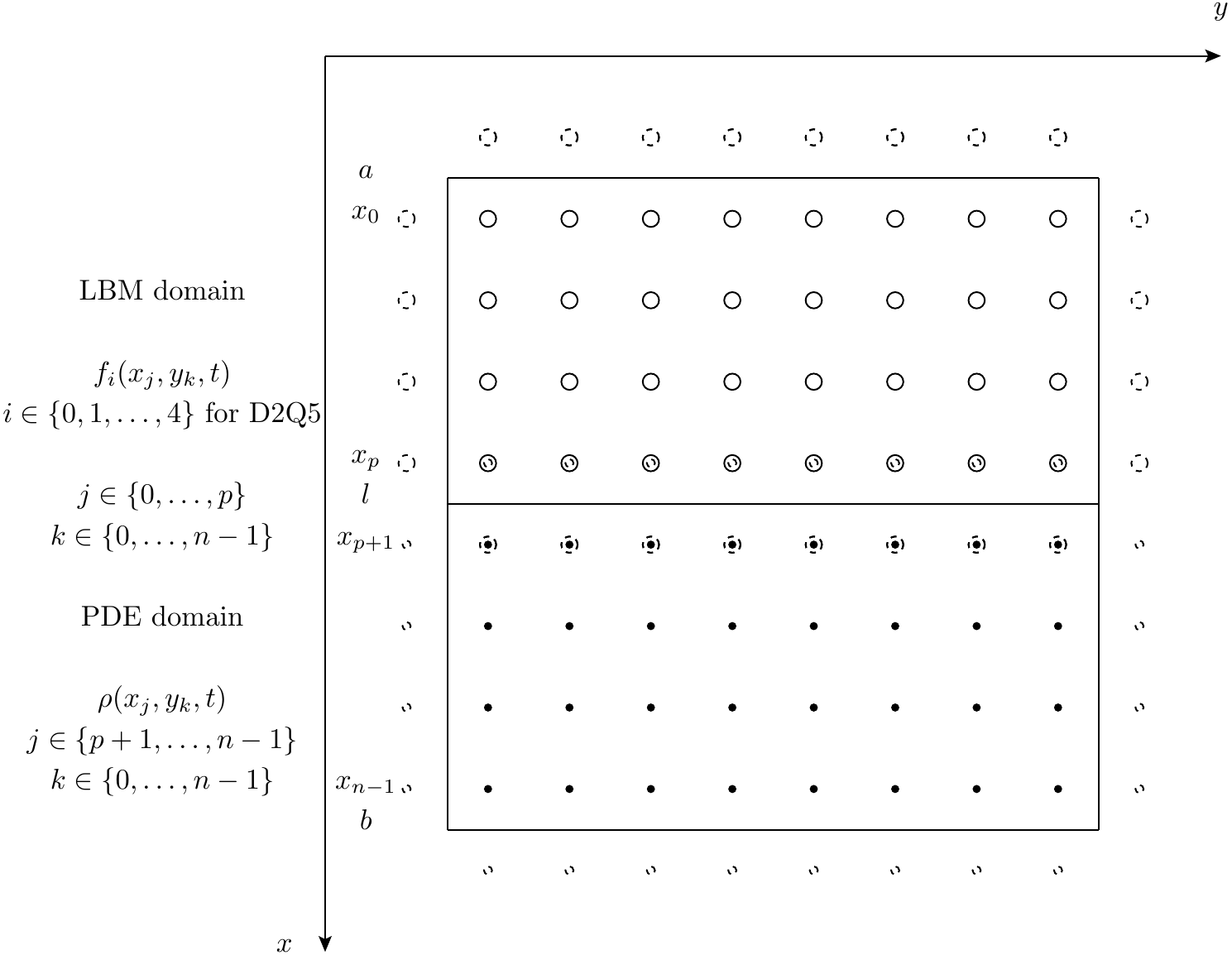}
\caption{The two-dimensional spatial domain $[a,b]\times[a,b] \subset \mathbb{R}^2$ in
  the hybrid model is split into
  $[a,l[\times[a,b]$ on which we solve the LBM and
  $[l,b]\times [a,b]$ on which we solve the PDE model.  The
  solid points ($\bullet$) represent the grid for the density $\rho$
  of the discrete PDE, the circles ($\circ$) represent the LBM
  variables $ f_i(x,y,t), i \in \{0,\ldots,4\}$ for D2Q5. The periodic boundaries and the coupling
  are implemented with ghostcells which are drawn by dashed
  circles. The density in the ghostcells of the PDE domain, in
  $(x_{p}, y_k)$, $k \in \{0,\ldots,n-1\}$  and $(x_n,y_k)$, are found by
  taking $\sum_i f_i$ in $(x_{p},y_j)$ and
  $(x_{0},y_k)$, respectively.  However, the ghostcells for
  the LBM domain, in $(x_{-1},y_k)$ and $(x_{p+1},y_k)$, require a lifting operator
  that lifts $\rho$ to the distribution functions in these
  points. \label{2Ddomain}}
\end{center}
\end{figure}

The hybrid test domain for this section is represented in Figure \ref{2Ddomain} for D2Q5. Again, the domain is split into subdomains.
One part of the domain is described by the LBM while another part is described by a macroscopic PDE. Example \ref{ex:param_D2Q5} describes the parameters for the model problem in this two-dimensional setting.

\begin{example} \label{ex:param_D2Q5} The considered model problem has the following parameters for a two-dimensional domain --- described by 5 possible velocity directions (D2Q5) --- of length $L \times L$ (with $n^2$ the number of grid points).
\[
L = 10, \quad n = 200, \quad \Delta x = \Delta y = \frac{L}{n}, \quad \Delta t = 0.0001, \quad \omega =1.6129.
\]
For these parameters the classical Chapman-Enskog expansion predicts a
diffusion coefficient $D=1$ (Eq.~\eqref{PDE_D2Q5}).
\end{example}

\begin{figure}[!htop]
\begin{center}
\includegraphics[width=0.45\textwidth]{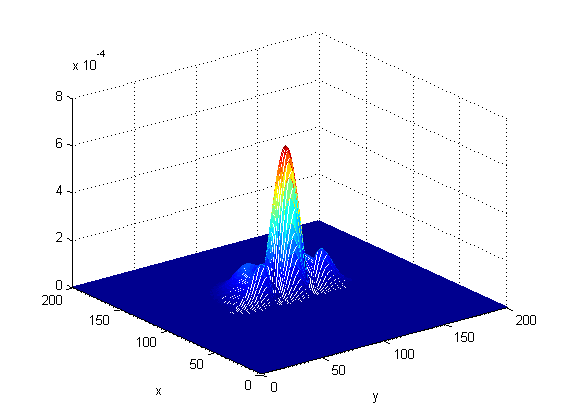} \includegraphics[width=0.45\textwidth]{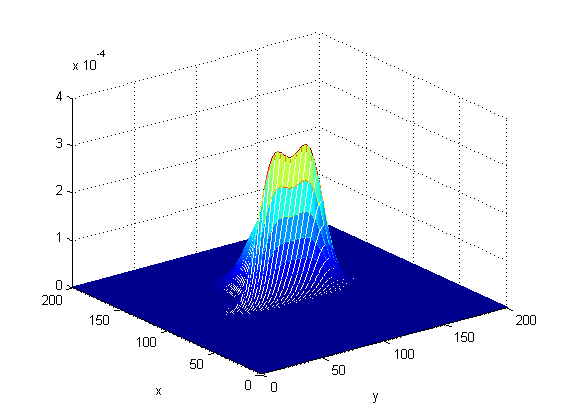}\\
\includegraphics[width=0.45\textwidth]{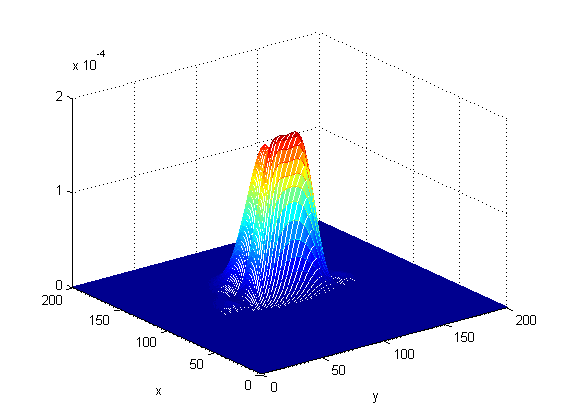}\includegraphics[width=0.45\textwidth]{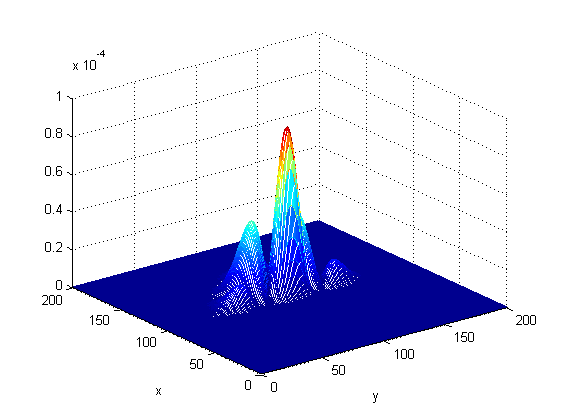}\\
\includegraphics[width=0.45\textwidth]{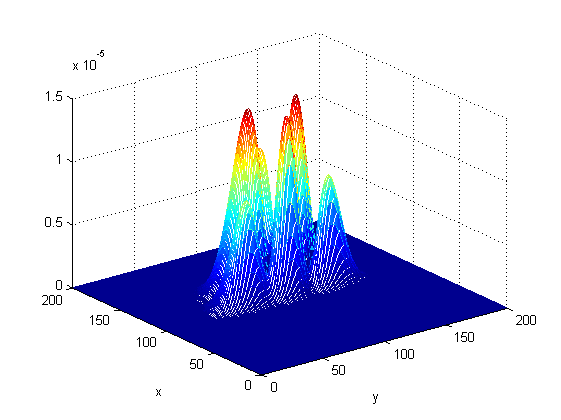}
\caption{
Comparison between different lifting operators for the model problem presented in Example \ref{ex:param_D2Q5}. The absolute difference $|\rho_{_{\text{hybrid}}}-\rho_{_{\text{LBM}}}|$ is plotted at time step 200. To
  deal with the initial error and the error in the ghost points of the
  LBM we use: the equilibrium distribution function (top left), the
  first order Chapman-Enskog expansion (top right), the second order
  Chapman-Enskog expansion (middle left), the numerical
  Chapman-Enskog expansion (order expansion 4) where the PDE in the hybrid
  domain is the analytically known PDE given in \eqref{PDE_D2Q5} (middle right) and 
  the numerical Chapman-Enskog expansion where the considered PDE in the hybrid domain is the one that is
  obtained from the numerical Chapman-Enskog expansion (bottom).
  \label{figure_D2Q5}}
\end{center}
\end{figure}

The comparison of $|\rho_{_{\text{hybrid}}}-\rho_{_{\text{LBM}}}|$ is represented in Figure \ref{figure_D2Q5} for Example \ref{ex:param_D2Q5}. The different lifting operators are used to obtain distribution functions from a given density. The used lifting operators are the equilibrium distribution function in the top left figure, the
  first order Chapman-Enskog expansion in the top right, the second order
  Chapman-Enskog expansion in the middle left, the numerical
  Chapman-Enskog expansion of spatial order expansion 4 in the middle right and the bottom --- depending on the used PDE in the hybrid model.

\subsubsection{D2Q9} \label{D2Q9}
This section takes more directions for the velocities into account. Example \ref{ex:param_D2Q9} contains the model problem parameters for D2Q9.
\begin{example} \label{ex:param_D2Q9} The considered model problem has the following parameters for a two-dimensional domain --- described by 9 possible velocity directions (D2Q9) --- of length $L \times L$ (with $n^2$ the number of grid points).
\[
L = 10, \quad n = 200, \quad \Delta x = \Delta y = \frac{L}{n}, \quad \Delta t = 0.00001, \quad \omega =1.9531.
\]
For these parameters the classical Chapman-Enskog expansion predicts a
diffusion coefficient $D=1$ (Eq.~\eqref{PDE_D2Q5}).
\end{example}

\begin{figure}[!htop]
\begin{center}
\includegraphics[width=0.45\textwidth]{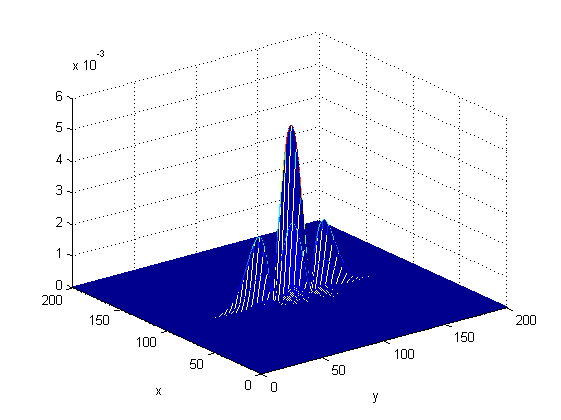} \includegraphics[width=0.45\textwidth]{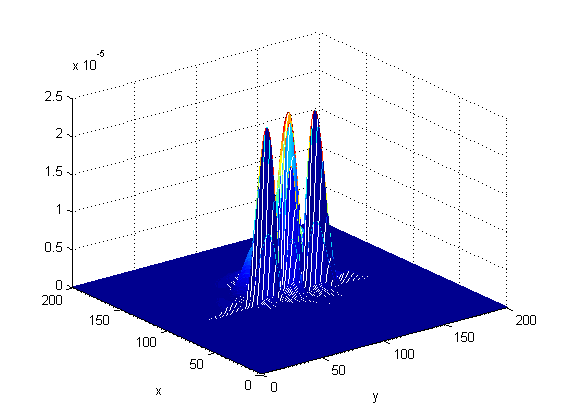}\\
\includegraphics[width=0.45\textwidth]{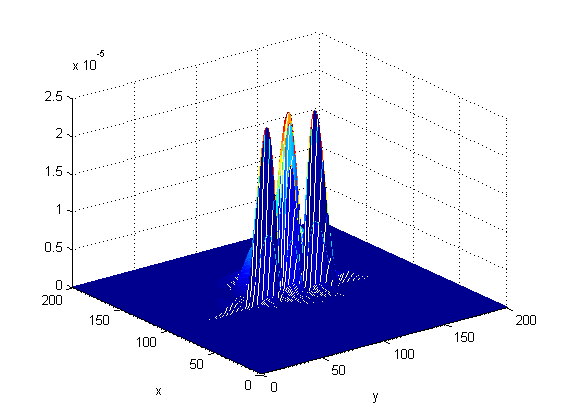} \includegraphics[width=0.45\textwidth]{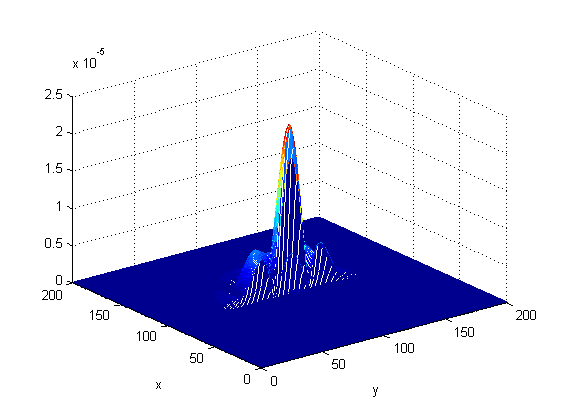}\\
\includegraphics[width=0.45\textwidth]{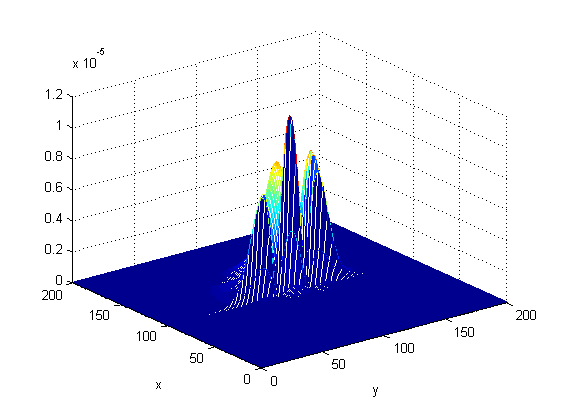}
\caption{$|\rho_{_{\text{hybrid}}}-\rho_{_{\text{LBM}}}|$ after 200 time steps with model problem presented in Example \ref{ex:param_D2Q9} without advection terms. To
  deal with the initial error and the error in the ghost points of the
  LBM the equilibrium distribution is used (top left), the first order Chapman-Enskog expansion (top right), the second order Chapman-Enskog expansion (middle left), the numerical
  Chapman-Enskog expansion (order expansion 4) where the PDE in the hybrid
  domain is the analytically known PDE given in \eqref{PDE_D2Q5} (middle right) and 
  the numerical Chapman-Enskog expansion where the considered PDE in the hybrid domain is the one that is
  obtained from the numerical Chapman-Enskog expansion (bottom).  \label{figure_D2Q9}}
\end{center}
\end{figure}

First consider no advection in the equilibrium distribution functions ($a=(0;0)$).
The results for $|\rho_{_{\text{hybrid}}}-\rho_{_{\text{LBM}}}|$ are presented in Figure \ref{figure_D2Q9}. The used lifting operators are the equilibrium distribution (top left), the first order Chapman-Enskog expansion (top right), the second order Chapman-Enskog expansion (middle left), the numerical
  Chapman-Enskog expansion (order expansion 4) where the PDE in the hybrid
  domain is the analytically known PDE given in \eqref{PDE_D2Q9} (middle right) and 
  the numerical Chapman-Enskog expansion where the considered PDE in the hybrid domain is the one that is
  obtained from the numerical Chapman-Enskog expansion (bottom).

\begin{figure}[!htop]
\begin{center}
\includegraphics[width=0.49\textwidth]{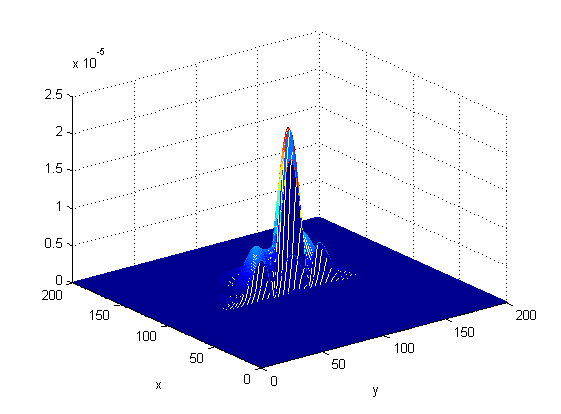}\includegraphics[width=0.49\textwidth]{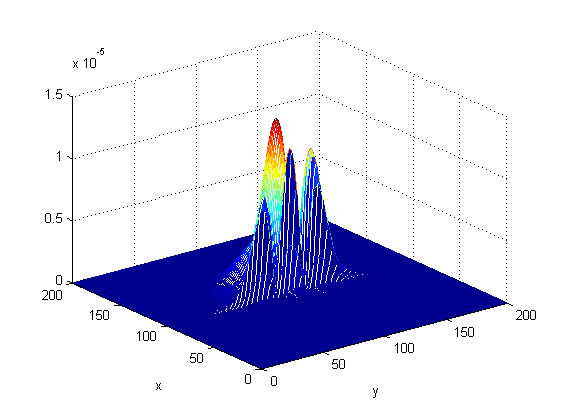}
\caption{$|\rho_{_{\text{hybrid}}}-\rho_{_{\text{LBM}}}|$ after 200 time steps with model problem presented in Example \ref{ex:param_D2Q9} with advection $a=(1;0.5)$. To
  deal with the initial error and the error in the ghost points of the
  LBM the numerical
  Chapman-Enskog expansion (order expansion 4) where the PDE in the hybrid
  domain is the analytically known PDE given in \eqref{PDE_D2Q9} (left) and 
  the numerical Chapman-Enskog expansion where the considered PDE in the hybrid domain is the one that is
  obtained from the numerical Chapman-Enskog expansion (right). \label{figure_D2Q9_advection}}
\end{center}
\end{figure}

When advection  ($a=(1;0.5)$) is included, we end up with Figure \ref{figure_D2Q9_advection} for the absolute difference $|\rho_{_{\text{hybrid}}}-\rho_{_{\text{LBM}}}|$ when the numerical Chapman-Enskog expansion is used to lift density to distribution functions.

\subsection{Analysis of the computational cost of lifting} \label{section_compare}
This section compares the computational cost of the lifting operators in the one-dimensional test problem (Section \ref{hybrid_1D}). 

The motivation for this paper is to bring down this cost. Especially the Constrained Runs algorithm requires many additional LBM steps to lift the density in the ghost points of the LBM domain. While numerical Chapman-Enskog only requires a single calculation with a fixed cost that can be done off-line before the simulation. This significantly reduces the cost of the lifting.

A detailed analysis of the lifting cost in terms of additional LBM steps is listed in Table \ref{compare_LBM}. The table is an
extension of the results of \cite{vanderhoydonc} with results for the classical Chapman-Enskog expansion, Constrained Runs algorithm combined with Newton's method and the numerical Chapman-Enskog expansion.

It can be seen that the total number of LBM steps for the CR-algorithm are listed per ghost point and per time step. The number for the
numerical Chapman-Enskog expansion is the total for the entire domain and at all time steps since the calculations for the coefficients are done off-line.

In two-dimensional problems the numerical Chapman-Enskog expansion still has a
limited computational cost. Only a few additional coefficients need to
be determined associated with the extra spatial derivatives.

Note that the computational cost of applying the numerical Chapman-Enskog lifting operator is the same as applying the analytical
Chapman-Enskog operator. For each grid point we need the derivatives of $\rho$, which can be calculated by finite differences using the
densities at neighboring grid points.
\begin{table}[!htop]
 \caption{
  Analysis of the computational cost of the lifting operators in terms of the additional 
LBM steps.  Both Constrained Runs (CR)  and the numerical Chapman-Enskog (NCE) require additional LBM steps 
    to lift $\rho$ to the distribution function $f$. The table shows these additional steps to construct the 
Jacobian operator for the Newton iteration.  The listed values for the CR-algorithm even are per ghost point and per time step. 
While those of the proposed lifting operator are the total number for the entire domain and for all time steps together 
since the vectors of constants needs to be determined only once and can be reused throughout the rest of the simulation. 
\label{compare_LBM}}
\begin{center} \footnotesize
\begin{tabular}{|l | l| l| l|}\hline
Lifting operator & Number of &  LBM steps  & Total number  \\
 & iterations &  to perform & of LBM steps \\
 & & one iteration &  \\ \hline
\em Exact Chapman-Enskog & / & / & 0 \\
& & & \\
\em CR-algorithm & & & \em per ghost point \\
\em type of extrapolation in time & & & \em per time step \\
Constant & 3  & $19 \times 1$ & 57  \\
Linear & 3  & $31 \times 2$ & 186  \\
Quadratic  & 3 & $43 \times 3$ & 387 \\
Cubic & 3  & $55 \times 4$ & 660 \\
& & & \\
\em Numerical Chapman-Enskog & & & \em for entire domain \\
\em with 18 unknowns  & & & \em and all time steps\\
\em $\{\alpha,\beta,\delta,\epsilon,\theta,\iota\}$ & & & \\
Constant & 3/(4) & 19 & 57+2=59\\
Linear & 3/(4) & $19 \times 2$ & 114+2=116\\
Quadratic  & 3/(4) & $19 \times 3$ & 171+2=173\\
Cubic & 3/(4) & $19 \times 4$ & 228+2=230\\ \hline
\end{tabular}
\end{center}
\end{table}

\section{Conclusions} \label{conclusion} 
This article proposes a numerical lifting operator for
lattice Boltzmann models (LBMs) that maps a given density to the
corresponding distribution functions.  This new lifting operator is
based on the Chapman-Enskog expansion that writes the missing
distribution functions as analytical series of the density and its derivatives. The
coefficients of this expansion are now determined through a numerical method,
in contrast to the original expansion where they are found
analytically. The numerical method is based on the Constrained Runs algorithm that relies 
on the attraction of the dynamics toward the slow manifold.

A systematic numerical comparison of the accuracy and the
computational cost between the analytical Chapman-Enskog expansion,
the Constrained Runs algorithm and the new lifting operator is
performed in this article.  The cheapest way to lift is with the
Chapman-Enskog expansion.  However, the analytical expressions are not
always available for the system of interest. An alternative numerical lifting operator is Constrained Runs (CR),
but its computational cost grows significantly with the order of
accuracy. It needs many additional LBM steps to find the missing distribution
functions.

The new result and the main focus of this paper is a numerical lifting
method that combines the ideas of Constrained Runs and the
Chapman-Enskog expansion. Instead of using Constrained Runs to find
for each grid point the missing moments, we use Constrained Runs to
find the unknown coefficients of the Chapman-Enskog expansion.  This
numerical lifting method has several advantages. First, it
significantly reduces the number of unknowns in the lifting step: we
only need to find the coefficients rather than the full state
$f(x,v,t)$. And secondly, it can be done off-line before the
calculations. Indeed, once the coefficients are found they can be
reused every time step and every grid point to realize the lifting, at
no significant additional cost.  A third advantage is that the
expansion gives, as a spin-off, the transport coefficients of the
macroscopic PDE.

The new lifting operator, the numerical Chapman-Enskog expansion, is then used in a hybrid domain that spatially
couples a macroscopic partial differential equation (PDE) with a
lattice Boltzmann model. This creates a missing data problem at
the interfaces  since the PDE model has too few
variables to provide the LBM with the correct boundary conditions.

The numerical Chapman-Enskog expansion deals with this mismatch in
variables. It maps the variables of the PDE model to those of the LBM.
We evaluate and compare various lifting operators. In particular, we
have focused on a simple LBM and PDE model discretized with equal grid
and time steps such that the error created by the coupling can be
highlighted.  The paper presents numerical results both for 1D and 2D
hybrid domains where part of the LBM domain is replaced by the
macroscopic PDE.  In both cases the error associated with the coupling
can be made smaller than the modeling error, related to the PDE
approximation of the LBM.

This paper reports on our initial efforts where we have focused
  on a simple model problem with several limiting assumptions.  In the
  model we have assumed an equilibrium distribution function that
  depends only on the local density, while in general it also depends
  on the local momentum and temperature.  This limitation can be
  easily alleviated by considering a Chapman-Enskog expansion with a
  more general equilibrium function.  

A further assumption is that we used the same time and space
  grid for the PDE and the LBM.  This choice was made to highlight the
  error made by the coupling mechanism, the ease of implementation and
  to eliminate the error due to the different discretizations.
  However, there is no reason to prohibit different grid and time
  spacings.  Extra care is then needed to interpolate between time and
  grid spacings.  In practice, the grid of the PDE can be further
  coarsened, depending on local discretization errors.  Ideally, the
  hybrid model is embedded in an adaptive mesh refinement simulation,
  where at the finest level a LBM is used.  

We have also kept the boundary between the LBM and the PDE domain
  fixed during the simulation at an arbitrary position.  In the
  future, this boundary should be moved adaptively using an accuracy
  requirement based on the properties of the lifting operator. 

For the model problem with periodic boundary conditions studied
  in this paper, the Chapman-Enskog expansion exists everywhere and we
  could in principle put the boundary between the PDE and the
  Boltzmann model at any location, provided that we lift accurately.
  For general Boltzmann models, with complicated collision integral
  operators, such a Chapman-Enskog expansion might not exist
  everywhere in the domain.  Then a hybrid model can be constructed
  where a PDE can replace the Boltzmann model only in the regions
  where the Chapman-Enskog expansion is known to exist. 

This situation appears in the modelling of laser ablation where a
  laser heats a surface that consequently melts and evaporates. The
  escaping plasma plume can be described by a Boltzmann
  equation. Close to the melting surface a complicated non-equilibrium
  situation appears where escaping particles evaporate but particles
  that impinge on the melted surface condensate.  There is no
  Chapman-Enskog expansion that can describe this situation close to
  the surface.  Only away from the surface the plasma reaches an
  equilibrium situation.  A hybrid model will then use a full
  Boltzmann model near melt while a reduced PDE model can be used away
  from the surface.

\section*{Acknowledgments}
This work is supported by research project \textit{Hybrid macroscopic
  and microscopic modelling of laser evaporation and expansion},
G.017008N, funded by `Fonds Wetenschappelijk Onderzoek' together with
an `ID-beurs' of the University of Antwerp.

\end{document}